\documentclass[%
reprint,
superscriptaddress,
nofootinbib,
 amsmath,amssymb,
 aps,
prb,
]{revtex4-2}
\usepackage[margin=2cm]{geometry}
\usepackage{amsbsy}
\usepackage{amsmath}
\usepackage{amssymb}
\usepackage{graphicx,xcolor}
\usepackage{textcomp}
\usepackage{hyperref}
\usepackage{array}
\usepackage{makecell, booktabs, mathtools}

\begin{document}
\title{Learning phases with Quantum Monte Carlo simulation cell}

\author{Amrita Ghosh$^{\ddag}$}
\email{amrita.ghosh.physics@gmail.com}
\affiliation{Center for Theory and Computation, National Tsing Hua University, Hsinchu 30013, Taiwan}
\author{Mugdha Sarkar$^{\ddag}$}
\email{mugdha.sarkar@gmail.com}
\thanks{Corresponding author}
\affiliation{Department of Physics, National Taiwan University, Taipei 10617, Taiwan}
\affiliation{Institute of Physics, National Yang Ming Chiao Tung University, Hsinchu 30010, Taiwan}
\author{Ying-Jer Kao}
\affiliation{Department of Physics, National Taiwan University, Taipei 10617, Taiwan}
\affiliation{Center for Theoretical Physics, National Taiwan University, Taipei, 10607, Taiwan}
\author{Pochung Chen}
\affiliation{Center for Theory and Computation, National Tsing Hua University, Hsinchu
30013, Taiwan}
\affiliation{Physics Division, National Center for Theoretical Sciences, Taipei 10617, Taiwan}
\affiliation{Department of Physics, National Tsing Hua University, Hsinchu 30013, Taiwan}

\begin{abstract}
We propose the use of the ``spin-opstring", derived from Stochastic Series Expansion Quantum Monte Carlo (QMC) simulations as machine learning (ML) input data. It offers a compact, memory-efficient representation of QMC simulation cells, combining the initial state with an operator string that encodes the state's evolution through imaginary time. Using supervised ML, we demonstrate the input's effectiveness in capturing both conventional and topological phase transitions, and in a regression task to predict non-local observables. 
We also demonstrate the capability of spin-opstring data in transfer learning by training models on one quantum system and successfully predicting on another, as well as showing that models trained on smaller system sizes generalize well to larger ones. Importantly, we illustrate a clear advantage of spin-opstring over conventional spin configurations in the accurate prediction of a quantum phase transition.
Finally, we show how the inherent structure of spin-opstring provides an elegant framework for the interpretability of ML predictions.
Using two state-of-the-art interpretability techniques, Layer-wise Relevance Propagation and SHapley Additive exPlanations, we show that the ML models learn and rely on physically meaningful features from the input data. Together, these findings establish the spin-opstring as a broadly-applicable and interpretable input format for ML in quantum many-body physics.

\end{abstract}

\maketitle

\begingroup
\renewcommand{\thefootnote}{\ddag}
\footnotetext{These authors contributed equally to this work.}
\endgroup

\section{Introduction}
In the past decade, due to their ability to analyze complex, high-dimensional data and uncover patterns that are challenging to detect using traditional methods, neural networks and machine learning (ML) techniques have become an integral part of condensed matter physics research (for a review, see for \textit{e.g.} \cite{dawid2023modernapplicationsmachinelearning,JOHNSTON2022100231,Bedolla_2021}). In particular, classifying phases and learning information about them, using supervised \cite{carrasquilla_machine_2017,Broecker2017} and unsupervised \cite{PhysRevB.94.195105,PhysRevE.95.062122,vanNieuwenburg2017,https://doi.org/10.48550/arxiv.1707.00663} methods, have been a central focus of the ML approach.
More recently, the community has increasingly sought to interpret the decisions of ML models, in order to overcome their inherent ``black box" nature (see \cite{wetzel2025interpretablemachinelearningphysics} for a review). While various kinds of input data, such as eigenstates \cite{doi:10.7566/JPSJ.85.123706}, correlation functions \cite{Shiina2020,Broecker2017}, entanglement spectra \cite{PhysRevLett.125.170603,vanNieuwenburg2017}, winding numbers \cite{https://doi.org/10.48550/arxiv.1707.00663} etc. are used to train these models, their choice often require existing knowledge about the properties of the system along with analytical or numerical computations. In line with the philosophy of artificial intelligence, ML models should ideally autonomously identify relevant features with minimal data preprocessing.

Raw spin configurations obtained as snapshots from Monte Carlo simulations have been successfully used as ML input in phase classification and related tasks \cite{carrasquilla_machine_2017,PhysRevB.94.195105}. However, they do not fully encode the quantum nature of the system which might be crucial for characterization of quantum phase transitions \cite{PhysRevB.97.045207}.
There have been proposals to use the raw $(d+1)$-dimensional configuration space generated by quantum Monte Carlo (QMC) methods to train ML models for a $d$-dimensional system, where the additional dimension corresponds to the imaginary time $\beta=1/T$ (with $T$ being the temperature). For increasing system sizes and low temperatures, which is necessary to study quantum phase transitions, the imaginary time direction increases rapidly leading to large simulation cells (usually $\mathcal{O}(10^2)$ Gigabytes and higher), which is too large to be effectively handled by machine learning algorithms. In Ref.~\cite{PhysRevX.7.031038,PhysRevE.97.013306}, the auxiliary field configurations obtained from determinant QMC simulations were utilized successfully as ML input, however, for only small systems at high temperatures. To address the challenge of large simulation cells, the authors of Ref.~\cite{PhysRevB.99.121104} proposed a lossy compression scheme for the $(d+1)$-dimensional space-time configurations generated by Stochastic Series Expansion QMC, to reduce their size to a manageable level and successfully used them as input for a convolutional neural network (CNN). However, the above-mentioned QMC algorithm provides an in-built lossless encoding scheme which we shall utilize in this work.

The Stochastic Series Expansion (SSE) algorithm, developed by Sandvik \cite{PhysRevB.43.5950,AWSandvik_1992, PhysRevB.56.11678}, is a powerful QMC method particularly well-suited for studying quantum spin systems and bosonic models. It is based on a Taylor series expansion of the partition function, the terms of which are then sampled stochastically. Using two update procedures, namely the diagonal and off-diagonal operator update, the SSE method generates simulation cells which can be represented by an allowed arrangement of operators (or operator strings) acting on an initial state. The operator string uniquely defines how the state propagates through the imaginary time and returns back to itself. The periodicity of the state in the imaginary time direction reflects the trace operation in the partition function. 

During an SSE QMC simulation, the full set of propagated states in a single update is not stored on the computer memory. Instead, only one propagated state and the operator string are stored for each QMC update, from which the remaining states can be generated on the fly. This approach substantially reduces memory requirements compared to other world-line QMC methods \cite{10.1063/1.3518900}, where entire space-time configurations are stored. It would therefore be interesting to ask whether this unique compact representation of the simulation cell can act as a meaningful input for ML tasks. In this work, we propose to use as ML input the initial state (or first configuration) of the simulation cell combined with the operator string from each QMC update, which we call in short the ``spin-opstring". In comparison to previous studies that have used QMC simulation cells as input, our choice of input preserves all the information contained in a simulation cell while being memory-efficient.

To demonstrate the effectiveness of the spin-opstring ML input for phase classification and regression tasks, we applied supervised machine learning techniques to two different models. The first model is the bosonic $t_2-V_1$ model, which, as reported in Ref.~\cite{PhysRevB.101.035147}, is the simplest model to realize a checkerboard supersolid phase on a two-dimensional square lattice. Using supervised ML with labeled spin-opstring input data generated by SSE QMC, we successfully reproduced the phase transitions of this model. Additionally, we performed a regression task with the ML model to predict the superfluid density, which is in close agreement to the QMC results. To further evaluate the performance of the spin-opstring data in the context of topological phase transitions, we applied supervised ML to the one-dimensional Su-Schrieffer-Heeger (SSH) model \cite{PhysRevLett.42.1698}. We found that despite the absence of an obvious order parameter, the spin-opstring data is able to capture the transition correctly and predicts the transition point accurately.

We further explored the application of spin-opstring data for transfer learning tasks. First, we were able to predict phase transitions in the interacting SSH model using a ML model trained in the non-interacting case.
Next, we applied our ML model trained on data from the $t_2-V_1$ model to successfully predict a phase transition in the $t_1-V_1$ model.
 Additionally, we showed that it is possible to generalize a ML model trained with spin-opstring data from a smaller system to larger systems. 

With the spin-opstring representation proving effective across various ML tasks, we investigate its potential advantages over other data formats, such as spin configurations. To explore this, we study a two-dimensional hard-core boson model that hosts a weak topological insulator phase. We show that spin-opstring outperforms spin-state due to the additional quantum information it encodes. Additionally, we demonstrate that it also surpasses the operator string alone, highlighting the necessity of combining both components into a single representation. 

Having established the usefulness of this input datatype, we turn our focus to the interpretation of the decision-making process in the trained ML models. 
It is important to mention here that the majority of studies dealing with interpretability put their emphasis on spin states as the input data \cite{PhysRevB.96.184410,PhysRevB.96.205146,Dawid_2020}. 
The suitability of alternative input representations in such analyses 
remain mostly unexplored
\cite{PhysRevResearch.2.023283,Cybinski_2025}.
To the best of our knowledge, our study marks the first instance of interpreting ML models trained on simulation cell data.
The structure of the spin-opstring, composed of a spin-state and a sequence of operators, provides an elegant framework for interpreting ML models.
To this end, we utilize two well-known interpretability methods, Layer-wise Relevance Propagation (LRP) \cite{10.1371/journal.pone.0130140} and SHapley Additive exPlanations (SHAP) \cite{NIPS2017_7062} to understand which features in the data are most important to the model’s predictions. We 
developed a strategy to extract meaningful insights, based on different physical operators, from the large-dimensional feature scores obtained directly from LRP and SHAP. In the classification, regression and transfer learning tasks, we find that ML models trained with spin-opstring data are able to identify key physical features from each phase, showcasing their ability to learn and utilize essential quantum characteristics.

The paper is organized in the following manner. In Sec.~\ref{proposal} we give a brief description of the SSE QMC algorithm and describe the spin-opstring input data. In Sec.~\ref{method}, we outline the input data format for machine learning, followed by a discussion on the neural network architecture and its training procedure. Next, we present the results obtained from the ML model in the context of phase classification and regression in Sec.~\ref{application}.
Sec.~\ref{further_app} explores further applications of the spin-opstring input in transfer learning. 
We compare the performance of using only the spin-state or the operator string with that of the combined spin-opstring in Sec.~\ref{spin_or_opstr}. 
Sec.~\ref{interpretability} focuses into understanding the model’s decision-making process through the use of established interpretability methods.
Finally, we conclude with a summary of our findings in Sec.~\ref{conclusion}.


\section{Proposal}\label{proposal}
In quantum statistical mechanics, the primary quantity of interest is the expectation value of any operator $O$ in a system governed by a Hamiltonian $H$ at temperature $T$,
\begin{align}
\langle O \rangle=\frac{1}{Z}\mathrm{Tr}\{Oe^{-\beta H}\}, Z=\mathrm{Tr}\{e^{-\beta H}\}\,,
\end{align}
where $\beta=1/T$ and all fundamental constants are set to unity. In the presence of non-commuting operators in a Hamiltonian, the calculation of the exponential function $e^{-\beta H}$ in the partition function becomes non-trivial. The starting point of the Stochastic Series Expansion (SSE) method is the expansion of the exponential operator in the partition function in a Taylor series \cite{PhysRevB.43.5950},
\begin{align}
    Z=\sum_\alpha\sum\limits_{n=0}^\infty\frac{\beta^n}{n!}\langle\alpha|(-H)^n|\alpha\rangle .
\end{align}
Here, the trace is replaced by a sum over a complete set of basis states $\{|\alpha\rangle\}$ of the Hamiltonian in question. Next, the Hamiltonian is decomposed into classes of elementary interactions as,
\begin{align}
    H=-\sum_{a,b}H_{a,b}\,,
\end{align}
where the index $a$ denotes the operator class, such as diagonal $(a=1)$ or off-diagonal $(a=2)$ in the chosen basis, while $b$ represents the bond index, which may correspond to nearest-neighbor, next-nearest-neighbor, or any other bond type. With this decomposition, the $n$th power of the Hamiltonian in the partition function can be written as a sum over all possible products, or ``strings", of the operators $H_{a,b}$ (denoted by $S_n)$ :
\begin{align}
    (-H)^n=\sum_{S_n}\prod\limits_{p=1}^{n} H_{a(p),b(p)}.
\end{align}
This results in the partition function
\begin{align}
Z=\sum_\alpha\sum\limits_{n=0}^\infty\sum_{S_n} \frac{\beta^n}{n!} \langle\alpha|\prod\limits_{p=1}^{n} H_{a(p),b(p)}|\alpha\rangle.
\end{align}
In the above expression, the summation over $n$ varies from $0$ to $\infty$, resulting in operator strings or ``opstrings" $S_n$ of varying length, which are numerically inconvenient to handle. To overcome this difficulty the Taylor expansion is truncated at some finite cutoff $M$, the value of which is dynamically chosen in the algorithm in such a way that the exponentially small truncation errors can be neglected. For the convenience in computation, $(M-n)$ identity operators, $H_{0,0}=I$, are added to a string of $n$ non-identity operators so that the length of all opstrings become equal. Now, $(M-n)$ identity operators can be inserted among the $M$ total operators in $\binom{M}{M-n}=\frac{M!}{(M-n)!n!}$ ways. To avoid over-counting due to the identical nature of the identity operators, we can divide the partition function by $\binom{M}{M-n}$, leading to the expression,
\begin{align}
    Z=\sum_{\alpha}\sum_{S_M} \frac{\beta^n(M-n)!}{M!}\langle\alpha|\prod\limits_{i=1}^M H_{a(i),b(i)}|\alpha\rangle.
\end{align}
Now, the terms in the above expression can be used as relative probabilities for Monte-Carlo sampling, provided they are positive. While the positive-definiteness of the diagonal terms are always guaranteed, which can be achieved by simply adding a constant, the same condition for the off-diagonal terms requires the lattice to be bipartite.

\begin{figure}[t]
\centering
\includegraphics[width=0.5\textwidth]{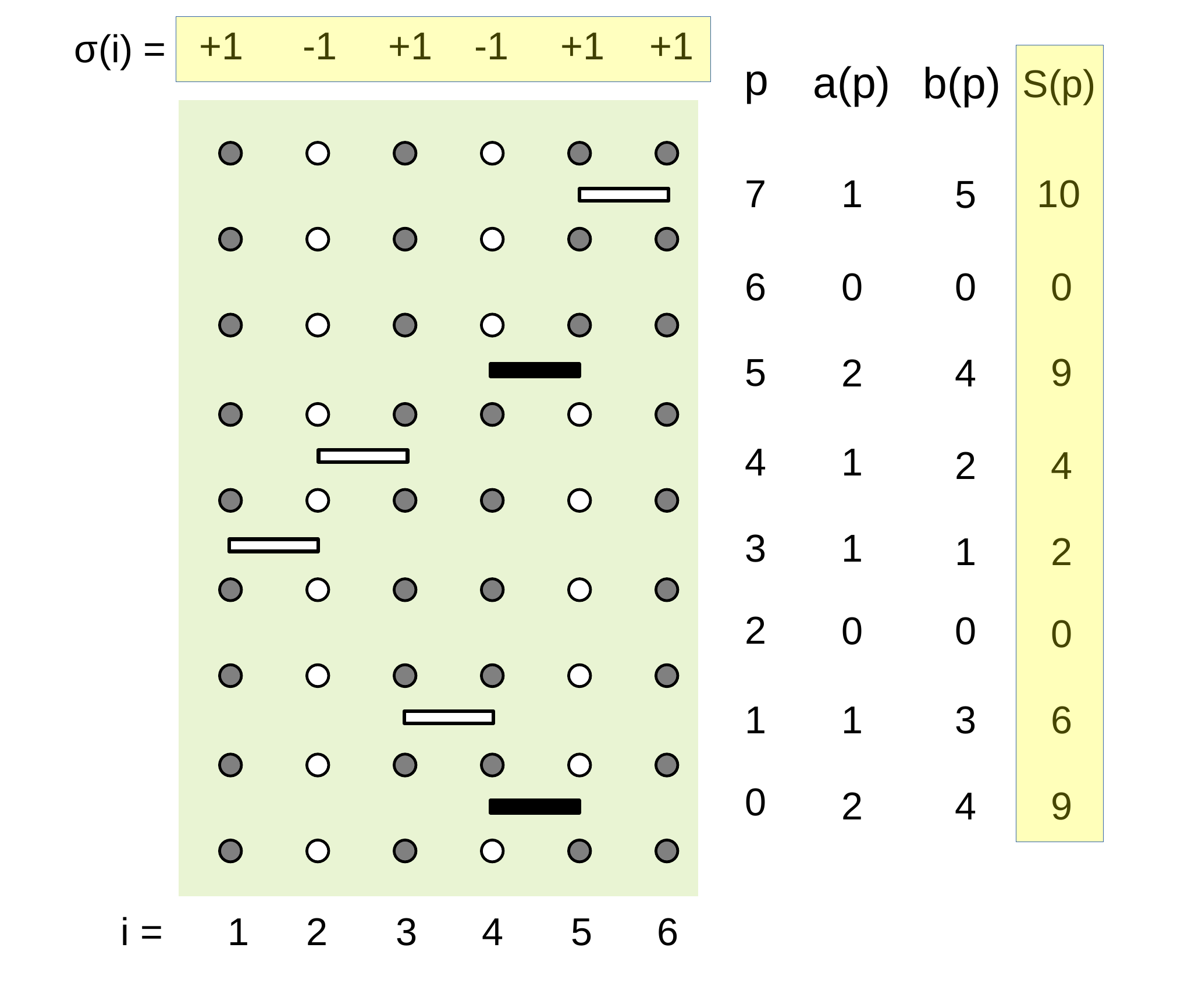}
\caption{Schematic diagram of the simulation cell of SSE QMC algorithm for a spin-$1/2$ system with six sites. Here $i$ signifies the lattice site number and $\sigma(i)$ represents the initial/final spin state. At each level $p$ of the simulation cell, the spin state is depicted by filled (+1)/open (-1) circles. The diagonal or off-diagonal operator at each level acting on a bond is represented by open or filled bars respectively. The information in the operator string is compactly codified in $S(p)$ ($a(p)$ and $b(p)$ are defined in the text). The ``spin-opstring" for the above simulation cell is constructed by combining $\sigma(i)$ and $S(p)$ together.}
\label{fig:sse_simcell}
\end{figure}

The update procedures in SSE QMC algorithm ensure sampling of states with non-zero contributions to the trace while preserving ergodicity. It is achieved by employing two types of updates,  diagonal update and off-diagonal update. The diagonal update adjusts the number of non-identity Hamiltonian operators $n$ in the operator string $S_M$, by replacing diagonal operators with identity ones and vice-versa. On the other hand, off-diagonal update replaces certain diagonal operators with off-diagonal ones and vice-versa, without altering the total number of operators $n$ in the operator strings. For a pedagogical introduction to the various SSE update procedures including recent developments, see \cite{Sandvik:2019jfn}.
It is important to note that, while SSE is fundamentally a finite-temperature QMC method, it can effectively capture a system's ground state when a sufficiently large inverse temperature $\beta$ is chosen \cite{PhysRevLett.74.2527}. For the systems examined in this study, we have observed that setting $\beta$ within the range of $1.5L$ to $2L$, with $L$ representing the linear dimension of the system, reliably led to the system converging to its ground state.

Fig.~\ref{fig:sse_simcell} depicts an example of a typical configuration or simulation cell generated by SSE QMC for a spin chain of $6$ sites. Here, the propagated states at a level $p$ (with $p=0,1,\cdots,M-1$),
\begin{align}
    |\alpha(p)\rangle=\prod\limits_{i=1}^p H_{a(i),b(i)}|\alpha\rangle,\quad |\alpha(0)\rangle=|\alpha\rangle,
\end{align}
are represented by solid and open circles corresponding to $\uparrow$ and $\downarrow$ spins, respectively, at the lattice sites. During an SSE simulation, the operator string $S_M$ along with one of the states $|\alpha(p)\rangle$ are stored. Therefore, in contrast to worldline simulations, where the whole space-time spin lattice is stored, the memory requirements for SSE are considerably smaller.

Taking advantage of this specific feature of SSE, in this paper, we propose to use the initial state $|\alpha(0)\rangle$ of the simulation cell in a given update, along with the operator string $S_M$ as the input data for ML models. In contrast to previous studies, where simulation cells have been used as input of ML models, our choice of input retains the entire information stored in a simulation cell, without being memory-extensive. 

In SSE simulation, the $\uparrow$ and $\downarrow$ spins are usually stored as $+1$ or $-1$, represented by $\sigma(i)$ in Fig.~\ref{fig:sse_simcell}, where $i$ indicates the site number. Moreover, in Fig.~\ref{fig:sse_simcell} the open and solid bars indicate diagonal and off-diagonal operators, respectively, while the absence of a bar between states correspond to a fill-in identity operator. The bond number on which an operator acts at level $p$ is denoted by $b(p)$. If a diagonal operator is present at level $p$, a value $1$ is assigned to the variable $a(p)$, whereas a value $2$ is assigned for an off-diagonal operator. Using this convention, an operator string $S_M$ can be constructed such that the diagonal and off-diagonal operators are encoded as even and odd integers, respectively, following the rule $S(p)=2b(p)+a(p)-1$, whereas for identity operators $S(p)$ is assigned to a value $0$. We append this operator string $S_M$ after the initial spin configuration and prepare the one-dimensional (1D) ``spin-opstring" input data. 

For a system with $N_s$ number of sites and an operator string of length M, the storage requirements for the total simulation cell is $N_s(1+M)$ memory units, whereas for the spin-opstring data it is just $(N_s+M)$. As an example for a $12\times 12$ square lattice ($N_s=144$) with a typical $M=25000$, the spin-opstring requires about $140$ times less memory than the whole simulation cell. In terms of actual numbers, for $30 000$ training samples of the two-dimensional (2D) $t_2-V_1$ model (discussed later in Sec.~\ref{sec:2dt2v}) on $12 \times 12$ lattices, the total size of the spin-opstring data is $\sim 5$ Gigabytes whereas for full simulation cells, it is around $700$ Gigabytes. Therefore, with increasing system sizes, the memory requirements for the total simulation cell data render them practically unusable, making the spin-opstring data a clear choice.

\section{Method}\label{method}
Before feeding the spin-opstring data to the machine, we preprocess the data by applying the following three steps. Firstly, the initial state $|\alpha\rangle$ is mapped from the spin basis to the occupation basis, such that $\uparrow$ spin at site $i$ ($\sigma(i)=1$) maps to the number operator $n_i=1$ and $\downarrow$ spin ($\sigma(i)=-1$) maps to $n_i=0$. Secondly, the operator string is normalized between values $0$ and $1$, by dividing the operator string elements by the maximum possible value, which is $2N_b+1$, where $N_b$ is the total number of bonds in the system. Lastly, since the length of the operator string varies in SSE QMC simulations at different regions of the phase diagram, we determine the maximum length of the training as well as prediction data and apply a padding of zeroes at the end of the operator strings as required to equalize the length of the data. As a zero in the operator string signifies the identity operator, padding with zeroes at the end does not change the physical properties of the system encoded in the data.

To reduce the correlation in the data, we collected spin-opstring data separated by several QMC steps. For example, the training data for phase classification in the $t_2-V_1$ model at each parameter value consists of 200 spin-opstring data picked from a QMC run of length $10^5$ MC steps. The phase classification tasks are achieved with the data being labeled by one-hot encoded vectors, while for regression, we provide the QMC expectation value of the observable as labels. During prediction, for each value of the tuning parameter, the ML model's output is averaged over multiple spin-opstring data sets obtained from a QMC run at the corresponding parameter value, for both classification and regression tasks. This mimics the idea of an expectation value of an observable for the case of regression. In the classification task we predict the phase boundaries by the point of maximum confusion in the model output, \textit{i.e.}, where the outputs of two softmax neurons intersect. 

To analyze the 1D spin-opstring data, we have used the usual feed-forward neural network architecture with fully-connected (FC) layers and/or 1D convolutional neural network (CNN) layers. Each of these layers consists of an initial fully connected or convolution step, followed by batch normalization and a rectified linear unit (ReLU) activation function. For the final layer, a fully connected layer is used with a softmax activation function for classification tasks, and a ReLU activation for regression tasks, without batch normalization in both cases. During training, we have minimized the categorical cross-entropy loss function for classification tasks and the mean squared error for regression. The stochastic gradient descent learning process has been optimized by using the Adam algorithm \cite{kingma2017adam}. The code has been implemented in Python by using the Keras library \cite{chollet2015keras}.

For the phase classification in the 2D $t_2-V_1$ model, we have used a FC ML model with three layers: the first input layer consisting of a number of neurons equal to the spin-opstring data length, followed by a hidden layer of 32 neurons, and finally an output layer consisting of a few neurons, depending on the number of phases, with softmax activation. In contrast, for regression, the ML model consists of two fully-connected hidden layers, each with $64$ neurons and the final layer has a single neuron with a ReLU activation function. 

\begin{table}[] 
\begin{tabular}{>{\centering\arraybackslash}m{2cm}|>{\centering\arraybackslash}m{2.4cm}|>{\centering\arraybackslash}m{2.2cm}} 
\multicolumn{2}{c|}{\makecell{Classification \\ model}} & \makecell{Regression \\ model}\\ \hline\hline
 \parbox{1cm}{Input layer}   & Conv1D  $f_n= 16$, $f_s= 9$ BatchNorm ReLU   & \parbox{1cm}{Input layer}\\ \cline{2-2}
   & Maxpool $s=2$    \\ \hline
 FC $n=32$ BatchNorm ReLU   & Conv1D  $f_n= 32$, $f_s= 5$ BatchNorm ReLU & FC $n=64$ BatchNorm ReLU   \\ \cline{2-2}
           & Maxpool $s=2$       \\ \cline{2-3} 
     & Conv1D  $f_n= 64$, $f_s= 5$ BatchNorm ReLU & FC $n=64$ BatchNorm ReLU  \\ \cline{2-2}
           & Maxpool $s=2$       \\ \cline{2-2}
     & Conv1D  $f_n= 128$, $f_s= 5$ BatchNorm ReLU   \\ \cline{2-2}
           & Maxpool $s=2$       \\
           \cline{2-2}
           & Flatten            \\ \hline
    \multicolumn{2}{c|} {\makecell{FC $n=2$ or $3$ \\ Softmax}} & \makecell{FC $n=1$ \\ ReLU}
 \\ \hline 
\end{tabular} 
\caption{Architectural setup of ML models used for classification and regression tasks. The abbreviations used above are as follows: FC: fully connected, $n$ : neuron number, $f_n$ : filter number, $f_s$ : filter size and $s$ : stride. In the first layers of the fully connected architectures, the number of neurons is equal to the size of the input spin-opstring data.} 
\label{tab:model-summary} 
\end{table}

For phase boundary detection in the 1D SSH model, we have used a CNN architecture. After the input layer, this architecture includes $4$ hidden CNN layers with $16, 32, 64$ and $128$ filters of sizes $9, 5, 5$ and $5$, respectively.
Each CNN layer is followed by a MaxPool layer with a pool size of $2$ and a stride of $2$, which reduces the length of the 1D data by half.
The output of the last CNN layer is then flattened and passed to a FC layer with two neurons, corresponding to the two phases of the SSH model. 
The ML models used for the phase classification and regression tasks discussed above are schematically represented in Table \ref{tab:model-summary}.

For both of the domain adaptation tasks,
we have employed 1D CNNs with a similar base architecture. Each network consists of three convolutional layers with kernel sizes of $9$, $5$, and $5$, respectively.
These layers are followed by 
batch normalization, ReLU activation, and 
a MaxPool layer with a pool size of $2$ and a stride of $2$. 
The model trained on $t_2-V_1$ uses $8$, $16$ and $32$ filters in the convolutional layers. 
The output from the last CNN is flattened and passed through a FC layer with $8$ neurons, followed by batch normalization, ReLU activation, which connects to a final FC layer of $2$ neurons with softmax activation. In contrast the model trained on the non-interacting SSH, employs $16$, $32$ and $64$ filters. Instead of flattening, after the final block, global average pooling is applied, followed by a FC layer consisting of $2$ neurons with softmax activation. During training in the SSH case, we employed
dropout regularization after each Conv1D layer with a rate of $0.25$ to achieve better generalization.

For each of the ML models used in this study, we present our results obtained as an average of the predictions from 100 independently trained instances of the same model.

\section{Applications : Phase classification and observable prediction}\label{application}

\subsection{Two-dimensional $t_2-V_1$ model}\label{sec:2dt2v}

 \begin{figure}[t]
\centering
\includegraphics[width=0.4\textwidth]{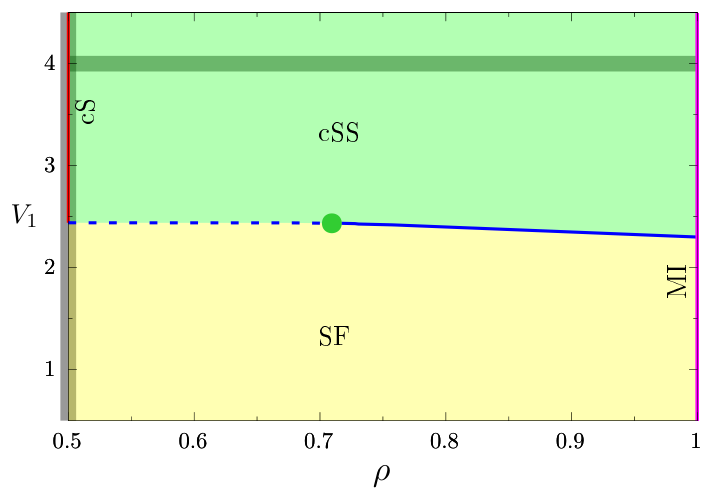}
\caption{Phase diagram of the $t_2-V_1$ model in Eq. \ref{Eq:Hamiltonian} in terms of particle-density $\rho$ and NN repulsion $V_1$, where the NNN hopping $t_2$ is set to unity. The dashed and solid blue lines represent mixed and continuous phase transitions, respectively, between the SF and cSS phases. The green filled circle denotes the tricritical point separating the mixed transition line from the continuous one. The shaded vertical and horizontal bands indicate the lines where we apply our ML model to predict the phase transitions.}
\label{fig:pdt2v}
\end{figure}

We first apply our proposal to the 2D $t_2-V_1$ model which belongs to the family of extended Bose-Hubbard models in the hardcore limit. This simple model displays a rich phase diagram and is, in fact, the minimum model to realize lattice supersolidity of the checkerboard-type. Interestingly, the system also exhibits a line of mixed-order phase transitions with varying particle density which displays the extreme Thouless effect at half-filling. In the 2D $t_2-V_1$ model, a system of hard-core bosons (HCBs) is studied on a periodic square lattice described by the Hamiltonian,
\begin{align}
    H=-t_2\sum_{\langle\langle i,j\rangle\rangle}(b^\dagger_i b_j+\textrm{h.c.}) 
    +V_1\sum_{\langle i,j\rangle}n_in_j-\mu\sum_i n_i,
    \label{Eq:Hamiltonian}
\end{align}
where $b^\dagger_i$ and $b_i$ are bosonic creation and annihilation operators and $n_i=b^\dagger_i b_i$ is the number operator at a site $i$. In this description, $\langle i,j\rangle$ and $\langle\langle i,j\rangle\rangle$ represent nearest-neighbor (NN) and next-nearest-neighbor (NNN) site pairs respectively. The hopping amplitude between NNN sites is denoted by $t_2$, the NN repulsion between the HCBs is represented by $V_1$ and $\mu$ indicates the chemical potential. 

In Ref.~\cite{PhysRevB.101.035147} this model was thoroughly investigated using SSE QMC technique by calculating two order parameters, structure factor and superfluid density. The structure factor per site is given as,
\begin{align}
    S(\boldsymbol{Q})=\frac{1}{N_s^2}\sum_{i,j}e^{i\boldsymbol{Q}\cdot(r_i-r_j)}\langle n_in_j\rangle,
\end{align}
where $r_i=(x_i,y_i)$ represents the position of site $i$, $N_s$ the total number of sites, $\boldsymbol{Q}$ the wave vector and $\langle\cdots\rangle$ denotes ensemble average. For the calculation of the superfluid density using SSE method, the following formula, which involves the winding numbers $W_x$ and $W_y$ along $x$ and $y$-directions, is used,
\begin{align}
    \rho_s=\frac{1}{2\beta}\langle W_x^2+W_y^2\rangle.
\end{align}

\begin{figure}[t]
    \centering
    \includegraphics[width=0.9\linewidth]{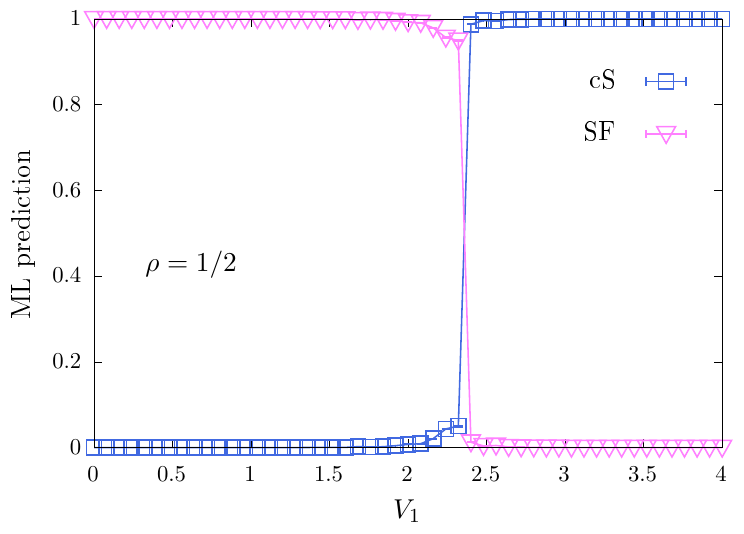}
    \includegraphics[width=0.9\linewidth]{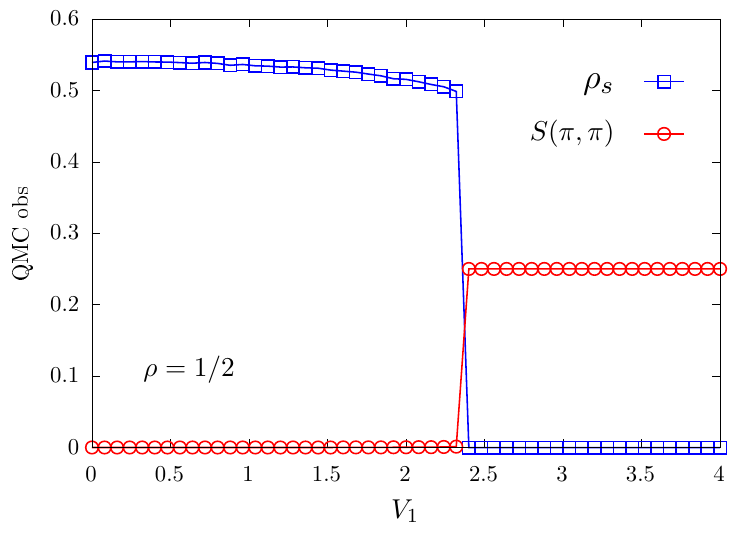}
    \caption{\textit{Top:} ML prediction for the two phases, cS and SF, as a function of the NN repulsion $V_1$ for a fixed particle density $\rho= 1/2$. \textit{Bottom}: QMC results for the observables, superfluid density $\rho_s$ and structure factor $S(\pi,\pi)$ as a function of $V_1$ for a $12\times12$ square lattice at half-filling.}
    \label{fig:VT_12x12}
\end{figure}

\begin{figure}[t]
    \centering
    \includegraphics[width=0.9\linewidth]{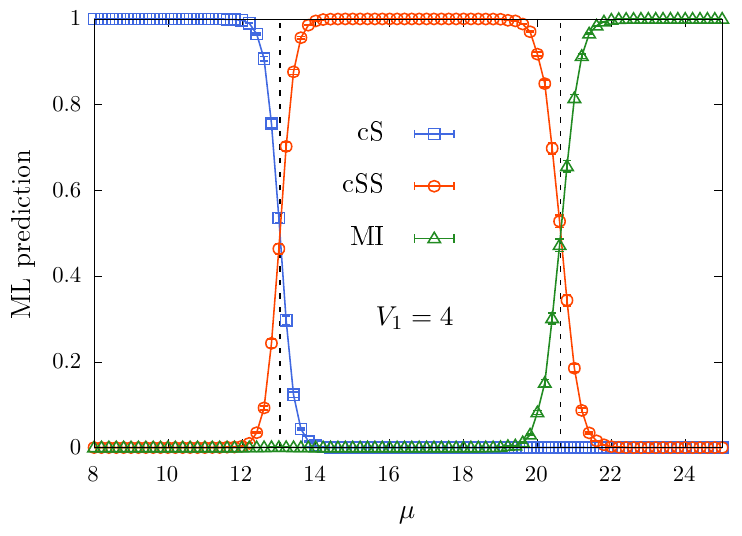}
    \includegraphics[width=0.9\linewidth]{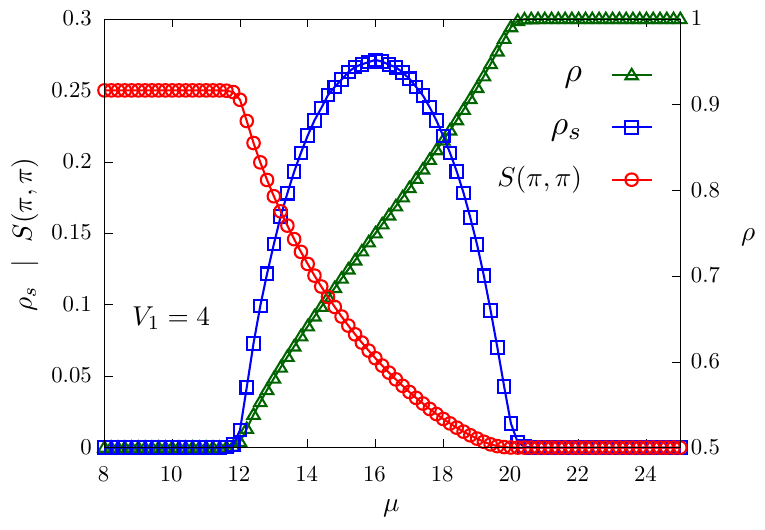}
    \caption{\textit{Top:} ML prediction for the three phases, cS, cSS and MI, as a function of the chemical potential $\mu$ for a fixed NN repulsion $V_1=4$. The vertical dashed lines indicate the ML predicted phase boundaries. \textit{Bottom}: QMC results for the observables, particle-density $\rho$, superfluid density $\rho_s$ and structure factor $S(\pi,\pi)$ with varying chemical potential $\mu$, on a $12\times12$ square lattice. The left $y$-axis shows the variation of $\rho_s$ and $S(\pi,\pi)$, while the right $y$-axis represents the variation of $\rho$.}
    \label{fig:HT_12x12}
\end{figure}

The study revealed that at half-filling, as the NN repulsion is increased beyond a critical value $V_1^c$, the system undergoes a mixed-order phase transition (displaying the extreme Thouless effect) from a superfluid (SF) phase to a checkerboard solid (cS) phase, where either one of the sub-lattices is completely occupied. For $V_1>V_1^c$ as the chemical potential $\mu$ is varied, the system goes through two continuous phase transitions. The first transition takes the system from the cS phase to a checkerboard supersolid (cSS) phase, where all the particles occupy a single sub-lattice and hop within it. At the second phase transition, the cSS phase becomes a Mott insulator (MI) at particle-density $\rho=1$, where all the lattice sites are completely filled. At intermediate particle densities between $\rho=1/2$ and $1$, the model displays either a mixed-order or continuous phase transition (separated by a tricritical point) between the SF and cSS phases as a function of $V_1$. In Fig.~\ref{fig:pdt2v} the above discussion is summarized in the phase diagram of the model, obtained from the SSE QMC study on a $24\times24$ square lattice, depicted in terms of particle-density $\rho$ and NN repulsion $V_1$. It should be noted that in the SSE QMC simulations, different values of particle-densities $\rho$ are accessed by tuning the chemical potential $\mu$ of the system. In particular, for the insulating phases, cS and MI, which are present at $\rho=1/2$ and $1$ respectively, one obtains plateaus in the particle-density $\rho$ as a function of $\mu$.

We apply a supervised ML model trained on the QMC spin-opstring data to predict the phase transitions. For the training, we collect $10000$ spin-opstring data in a small window from deep within each of the phases : cS, cSS, SF and MI\footnote{In particular for cS and MI phases, deep within the phase precisely means the $\mu$ values around center of the plateau in $\rho$.}. The trained ML model is then used to predict unlabeled data along two lines in the phase diagram: firstly, the vertical line at half-filling,  
as a function of $V_1$, and secondly, the horizontal line at a fixed $V_1=4.0$ as function of $\mu$, and hence as function of particle-density $\rho$.
These two lines are indicated by shaded grey bands in Fig.~\ref{fig:pdt2v}.
Notably, we observe considerable variation in opstring lengths across phases, with the longest lengths occurring in the MI phase and the shortest in the SF phase. 

\begin{figure}[t]
    \centering
    \includegraphics[width=0.8\linewidth]{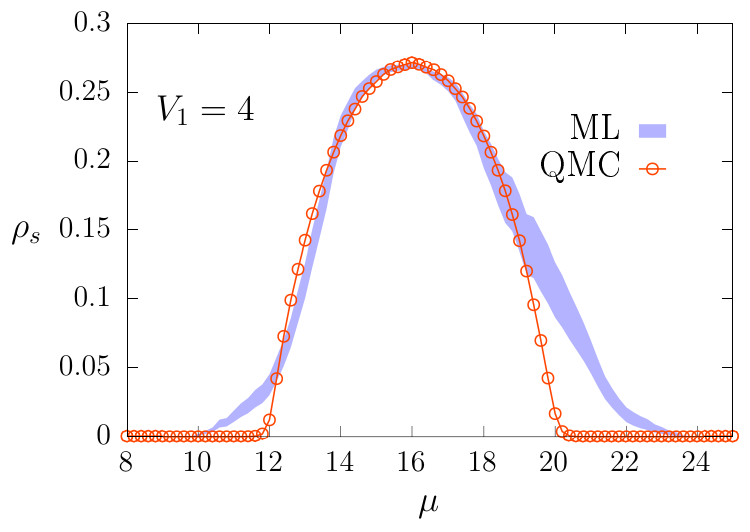}
    \caption{Comparison of superfluid density obtained from the ML regression model with QMC results on a $12\times 12$ square lattice. The error band of the ML data has been obtained from 100 independently trained ML models.}
    \label{fig:HT_reg}
\end{figure}  
In Fig.~\ref{fig:VT_12x12} top, we show the ML prediction
on 
a $12\times 12$ square lattice at half-filling with a varying NN repulsion $V_1$. 
A sharp transition between the SF and cS phase is displayed by a jump between the extremum probability values of the output neurons. We contrast the ML prediction with the QMC results on the same lattice size (Fig.~\ref{fig:VT_12x12} bottom), where superfluid density $\rho_s$ and structure factor $S(\pi,\pi)$, are plotted with increasing $V_1$. For $V_1$ below a critical value, 
the presence of the SF phase is characterized by a near-maximal value of $\rho_s$ and a zero $S(\pi,\pi)$. With increasing $V_1$, the system transitions to the cS phase, where $S(\pi,\pi)$ reaches its maximum value, and $\rho_s$ drops to zero. This sharp jump in the order parameters to their respective extrema values is a signature of the extreme Thouless effect and this behavior is also reflected in the ML predictions. Notably, the transition point predicted by the ML model matches perfectly with the QMC result, identifying the critical NN repulsion as $V_1=2.36\pm 0.04$. It is important to mention here that we found the Batch Normalization layer to be a crucial ingredient in the ML model for making consistent predictions. We provide details regarding this in Appendix \ref{app1}.

Next, in Fig.~\ref{fig:HT_12x12} top, we present the ML predictions for the same system size at fixed $V_1=4.0$ as a function of $\mu$, revealing two phase transitions, from cS to cSS, and from cSS to MI, both of which appear to be continuous. 
In the bottom figure, we show the same phase transitions demonstrated by the QMC observables $\rho_s$ and $S(\pi,\pi)$, along with the particle-density $\rho$ as a function of $\mu$. The figure illustrates the existence of three phases: checkerboard solid at density $\rho=1/2$ where only the structure factor $S(\pi,\pi)$ is non-zero, checkerboard supersolid for intermediate densities where both $S(\pi,\pi)$ and $\rho_s$ are non-zero and the Mott insulator at density $\rho=1$ where both the observables, $S(\pi,\pi)$ and $\rho_s$, are zero.
In the phase diagram in Fig.~\ref{fig:pdt2v}, these two transitions are depicted in terms of the corresponding particle density $\rho$ highlighted by the horizontal shaded band. In this case as well the ML predictions match well with the QMC results.  

Next, to evaluate the performance of the spin-opstring data type in regression tasks, we selected superfluid density $\rho_s$, a non-local observable connected to the winding of particles through the spatial dimensions of the system. 
The regression task is performed along the same line as in Fig.~\ref{fig:HT_12x12}, with $V_1=4.0$, using the same data, now labeled with the values of $\rho_s$. Since $\rho_s$ is an expectation value calculated by averaging over several simulation cells, all samples at a given $\mu$ share the same $\rho_s$ value as label. Similarly, ML predictions are averaged over multiple samples for each $\mu$ to obtain the final prediction for $\rho_s$.  
In Fig.~\ref{fig:HT_reg}, we compare the prediction of the regression model with the QMC results obtained for a $12\times12$ square lattice. For the most part the predicted superfluid density values closely agree with the QMC results.
However, near the transition points unsurprisingly the machine prediction show deviations as it encounters critical behavior which is absent in the training data. The regression results further support the suitability of our proposed input data for machine learning applications.

\begin{figure}[t]
    \centering
    \includegraphics[width=0.8\linewidth]{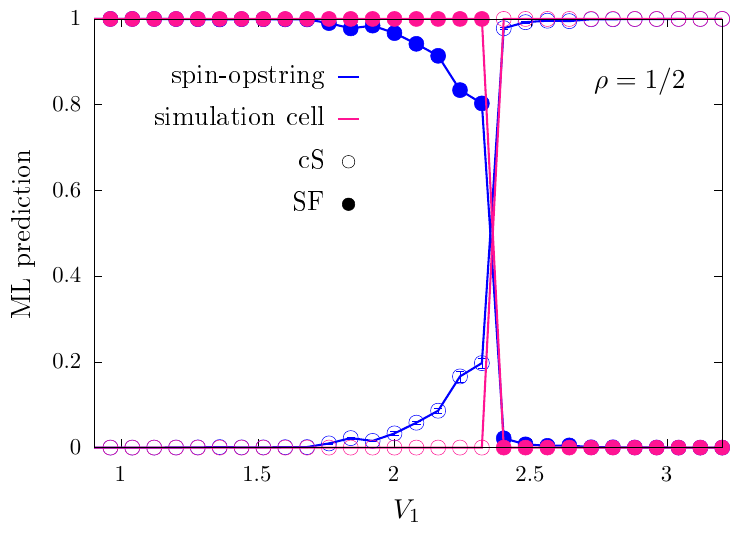}
    \caption{Comparison of ML models trained with spin-opstring and full simulation cell data for the SF-cS transition on $8\times 8$ lattices. The open symbols denote the probability for cS phase while the filled symbols stand for the SF phase. The errors on the points have been obtained from 20 independent training setups.}
    \label{fig:simcellcomp}
\end{figure}

Although the spin-opstring data provides a memory-efficient compact representation of the full QMC simulation cell, it would be instructive at this stage to do a comparison of both of them. For this purpose, we consider the SF-cS transition on a $8\times 8$ lattice. 
We chose this smaller lattice because the size of the full simulation cell remains within feasible computational limits (each spin-opstring data has $64+5282=5346$ elements while each full simulation cell has $64\times (5282+1)=338112$ entries).
Unlike in the case of spin-opstring, the use of a FC ML architecture is not viable for the full simulation cell as it leads to $\mathcal{O}(10^7)$ trainable parameters even for this smaller lattice.
We use instead a 2D CNN network leveraging the natural two-dimensional structure of the simulation cell where the horizontal direction is the site index of the spin-state and the vertical direction represents its time evolution.

The 2D CNN model has a similar architecture as described for the 1D SSH case in Table \ref{tab:model-summary}, having only three CNN layers with $8, 16$ and $32$ filters respectively, followed by a FC classification layer. In Fig.~\ref{fig:simcellcomp}, we show the predictions for the phase classification obtained from the two different ML models trained with the same number of data points. Both of them predict exactly the same transition point which perfectly matches with the QMC result. 
It appears that the 2D CNN model trained with the full simulation cell data is able to better reflect the sharp transition associated with the extreme Thouless effect. However, due to the larger size of the full simulation cell data, the training time per epoch on a Nvidia A100 GPU is about $6$ seconds whereas for the spin-opstring data, it is a mere $80$ milliseconds, a $75\times$ speedup.

\begin{figure}[t]
    \centering
    \includegraphics[width=0.7\linewidth]{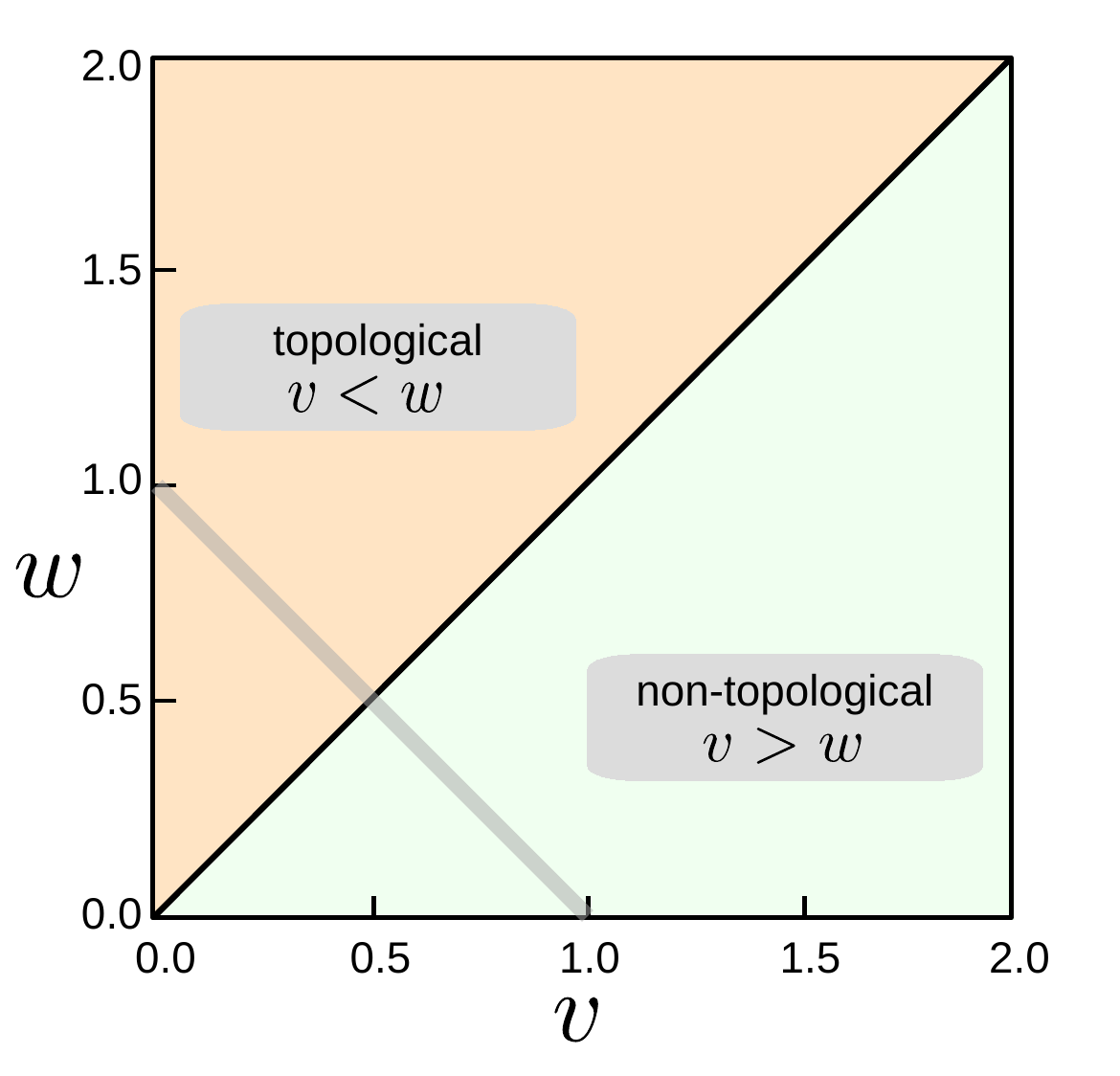}
    \caption{Phase diagram of 1D SSH model in terms of hopping amplitudes $v$ and $w$. The shaded gray band represents the line along which we apply our ML model to predict the phase transition.}
    \label{fig:ssh_PD}
\end{figure}
\begin{figure}[b]
    \centering
    \includegraphics[width=0.8\linewidth]{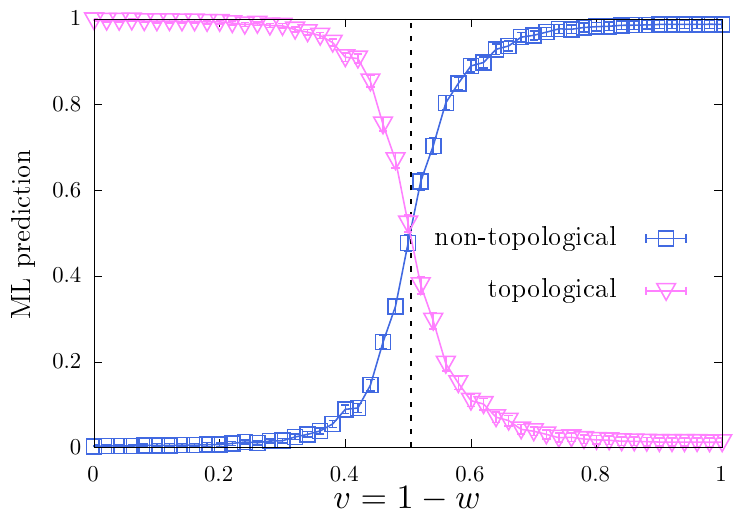}
    \caption{ML prediction of the SSH model performed on 
    a 1D periodic chain with $32$ sites. As $v$ is varied from $0$ to $1$, $w$ changes according to $1-v$. The dashed line represents the phase boundary predicted by the ML model.}
    \label{fig:sshml}
\end{figure}

\subsection{One-dimensional Su-Schrieffer-Heeger (SSH) model}
In the previous section we evaluated the spin-opstring data on the $t_2-V_1$ model which features conventional phase transitions driven by symmetry breaking and defined by order parameters. However, it would be intriguing to explore how this spin-opstring input data performs in the context of topological phase transitions. Unlike conventional transitions, topological transitions do not involve any local order parameter, but are instead characterized by changes in global topological properties of the system, such as the winding number or Chern number. 

In order to demonstrate the performance of the spin-opstring input data on topological phase transitions, we select the simplest and most fundamental model for topological excitations: the one-dimensional Su-Schrieffer-Heeger (SSH) model. 
While the SSH model is well understood and may seem unnecessary for SSE QMC simulations, it serves as 
a proof-of-concept, that helps demonstrating the effectiveness of spin-opstring data in the identification of topological phase transitions. 
It becomes particularly important later, in Sec.~\ref{trans_learning}, where we discuss the application of a ML model trained in the non-interacting SSH model to predict phase transitions in a interacting version, a scenario where QMC simulations become essential.

Originally, the SSH model was proposed to describe the behavior of electrons and the emergence of solitonic defects in a polyacetylene molecule \cite{PhysRevLett.42.1698}. It is a tight-binding model that describes spinless electrons on a one-dimensional dimerized chain, arising from alternating strong and weak hopping amplitudes between the nearest-neighbor sites. The Hamiltonian of the 1D SSH model is described as,
\begin{align}
    H=v\sum\limits_{j\in \text{odd}} c^\dagger_{j} c_{j+1} + w \sum\limits_{j\in \text{even}} c^\dagger_{j} c_{j+1}+ \textrm{h.c.}, \label{H_SSH}
\end{align}
where $c^\dagger$ and $c$ denote the electron creation and annihilation operators, respectively. The summations in the first and second term span over the odd and even lattice sites, respectively. The parameters $v$ and $w$ correspond to the hopping amplitudes along alternate bonds. As a result of the competition between these hopping amplitudes, the SSH model undergoes a topological phase transition at half-filling. 
A non-topological dimer phase ($v>w$) transitions into a topological phase ($v<w$) at the critical point $v=w$, for a fixed choice of boundary.
The phase diagram of 1D SSH model in terms of the hopping amplitudes $v$ and $w$, are presented in Fig.~\ref{fig:ssh_PD}.

To train the ML model for 1D SSH model, we collect 40000 spin-opstring data in a small window around each of the two points $(v,w)=(0.9,0.1)$ and $(0.1,0.9)$ which are symmetrically apart from the phase transition at $(0.5,0.5)$. These data are labeled ``topological'' and ``non-topological'' respectively in the training process. It should be noted that we have used periodic boundary conditions (PBC) while collecting the training and prediction data. Application of open boundary conditions (OBC) would lead to the appearance of edge states. Using spin-opstring data derived from QMC simulations under such boundary conditions would not therefore be in line with our aim of learning topological phase transitions in the absence of any local order parameter.

 The trained model is now fed unseen spin-opstring data along a line in the coupling space characterized by the equation $v=1-w$ for $v\in[0,1]$ which goes through the above-mentioned phase transition point. This line is indicated by the oblique shaded band in the phase diagram in Fig.~\ref{fig:ssh_PD}. The predictions of the ML model is shown in Fig.~\ref{fig:sshml}, where we can see that it successfully predicts the true phase transition at $(0.5,0.5)$ based on the spin-opstring data. This demonstrates the spin-opstring input data's capability to capture topological phase transitions that are independent of any local order parameter.
 
\section{Transfer Learning}\label{further_app}

In this section we extend our investigations to explore the capabilities of the spin-opstring data in transfer learning.
In the context of phase classification tasks, we consider the two following cases of transfer learning : i) a ML model trained on one system used to predict another related system, which is also known as \textit{domain adaptation}, and ii) a ML model trained on a system of given size used to predict the same system with a larger size.
These studies allow us to investigate whether the ML models capture universal properties from the input data rather than being dependent on non-generalizable traits.

\subsection{Domain Adaptation}\label{trans_learning}
For this purpose we consider the 1D SSH model.
We leverage the spin-opstring data to transfer knowledge from the non-interacting SSH model to its interacting version.
In the presence of NN repulsion, the Hamiltonian of the interacting SSH model reads as,
\begin{align}
H=H_{\mathrm{NI}} +V_1\sum_{\langle i,j\rangle} n_i n_j,
\end{align}
where $H_{\mathrm{NI}}$ corresponds to the non-interacting SSH Hamiltonian given in Eq.~\ref{H_SSH} and $\langle i,j\rangle$ denotes NN pairs of sites. Even with the inclusion of NN repulsion, the topological phase of the SSH model persists up to a certain interaction strength, with the transition to the trivial phase occurring at $v=w$. However, for larger values of $V_1$, a charge-density-wave (CDW) state emerges between the topological and trivial phases\cite{PhysRevB.27.1680,Sirker_2014,PhysRevB.108.195151}.

\begin{figure}
    \centering
    \includegraphics[width=0.75\linewidth]{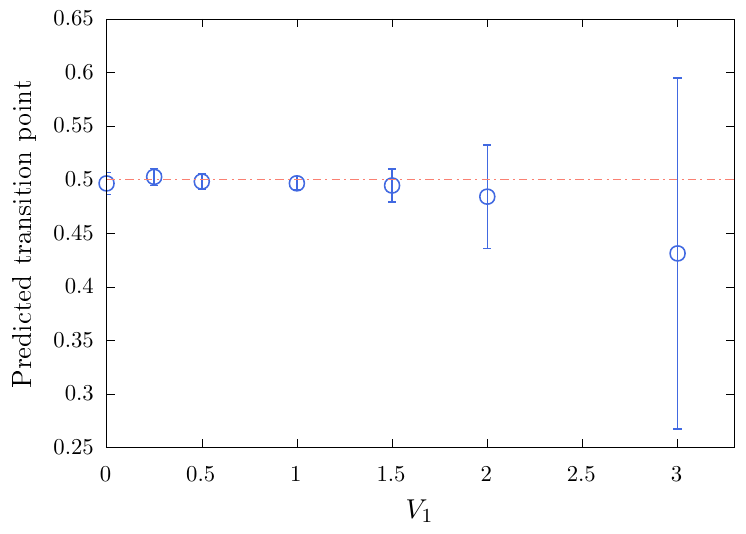}
    \caption{ML predicted transition point between the topological and trivial phases for interacting SSH model, in terms of NN repulsion $V_1$.}
    \label{fig:ssh_transfer}
\end{figure}

In our transfer learning setup, we trained a CNN model exclusively on data from the non-interacting SSH regime and tested its performance on the interacting SSH model. Fig.~\ref{fig:ssh_transfer} shows the transition points between the topological and trivial phases predicted by the ML model, as a function of increasing repulsion $V_1$. It demonstrates that the model is capable of transferring the knowledge learned in the non-interacting limit to accurately predict the transition points for the interacting system up to $V_1\approx 2$. The model delivers inconsistent predictions for larger repulsions as the distribution of states differs significantly from the non-interacting system, in particular, due to the emergence of the CDW phase for large $V_1$.

An additional example of domain adaptation, where a ML model trained on data from the $t_2-V_1$ model successfully predicts a phase transition in the $t_1-V_1$ model, is provided in Appendix~\ref{app2}.

In the context of domain adaptation in 1D SSH model, Ref.~\cite{10.21468/SciPostPhys.11.3.073} employed a supervised ML algorithm to identify topological phases in interacting systems by training a neural network on their non-interacting versions, with the curvature function at the high symmetry points as input. Another study \cite{PhysRevB.97.134109} also demonstrated
transfer learning in the SSH model, from a clean to disordered system, using eigenstates as input with domain adversarial neural networks.

\subsection{Generalization across system sizes}

Next, we wanted to investigate how well the spin-opstring performs when we train the ML model on a smaller system size and predict on a larger one. 
This capability is particularly useful, because simulating larger systems is often computationally expensive and time-consuming.
Moreover, since the length of the spin-opstring data increases with system size, using smaller system data for training would vastly reduce the  computational cost.
Therefore, the ability to generalize to larger system sizes is highly valuable, as it can significantly minimize computational time and resources.
\begin{figure}[t]
    \centering
    \includegraphics[width=0.85\linewidth]{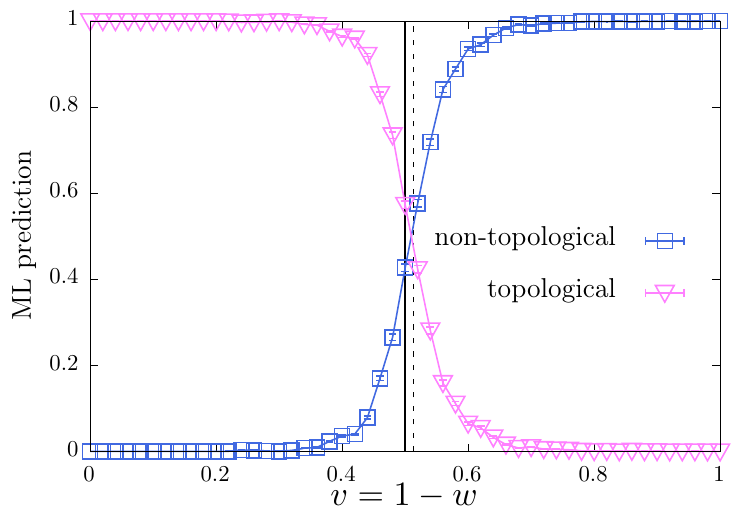}
    \caption{Prediction for the phase transition in the 1D SSH model with $32$ sites, using a ML network trained on data from a $24$-site lattice. The black vertical line denotes the exact transition point while the dashed one represents the predicted transition point.}
    \label{fig:largesystem}
\end{figure}

For this purpose, to handle the variable length data, a Global Pooling Layer was incorporated into our 1D CNN architecture (further details can be found in Appendix \ref{app3}.)
In Fig.~\ref{fig:largesystem} we show the prediction of the ML model when trained on the data generated for a $24$-site periodic SSH chain and tested on a system with $32$ sites. We can see that the model almost accurately predicts the transition point with less than $3\%$ deviation. This demonstrates that our network is not rigidly tied to the specific length of the training data, but rather extracts scale-invariant patterns of the phases.

\section{Spin-state or \textit{opstring} ?}\label{spin_or_opstr}

\begin{figure}[t]
    \centering
    \includegraphics[width=0.5\linewidth]{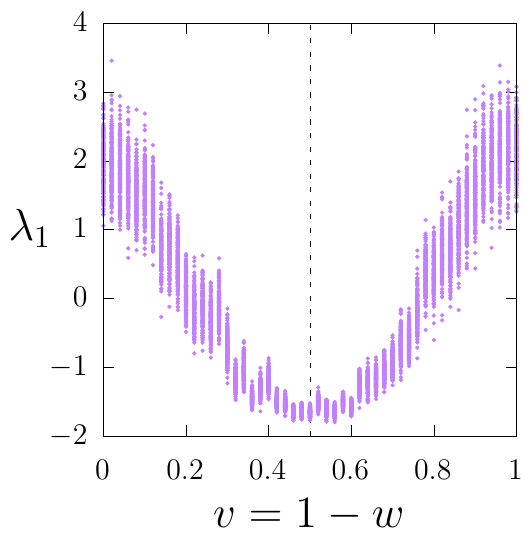}
    \includegraphics[width=0.46\linewidth]{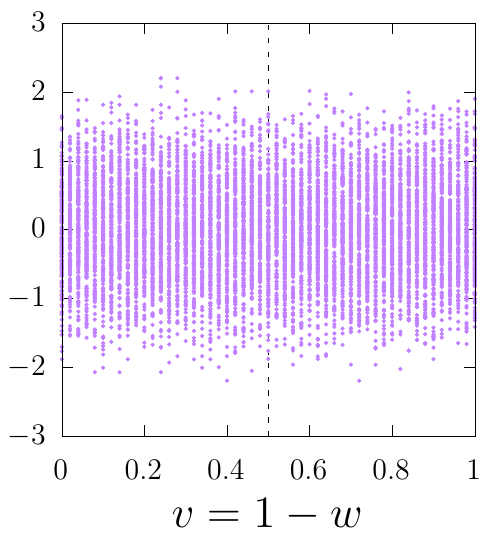}
    \caption{Variance $\lambda_1$ along the first principal component obtained from PCA done on spin-opstring (\textit{Left}) and spin-state (\textit{Right}) data independently for values of coupling across the transition in the $1D$ SSH model.}
    \label{fig:pca_ssh}
\end{figure}
 
The operator string combined with the initial spin configuration, equivalent to an entire QMC simulation cell, embodies quantum features unique to a particular phase. We have seen above that this data representation obtained from QMC simulations can be successfully used for classification, regression and transfer learning tasks. However, it has been demonstrated in previous studies that the spin configuration alone can prove successful for the above ML tasks in various quantum systems. 
Therefore, it is important to ask whether there exist scenarios where the additional quantum information encoded in the spin-opstring becomes crucial for accurate phase recognition. In this section, we provide an example of a 2D HCB system where the spin-state alone turns out to be insufficient.

Before proceeding with the example, let us consider the spin-state and spin-opstring data in the already discussed 1D SSH model.
We show the results of a principal component analysis (PCA) performed on the spin-state data in Fig.~\ref{fig:pca_ssh} right and on the spin-opstring data in Fig.~\ref{fig:pca_ssh} left.
The principal components (the variance of the first one is shown in Fig.~\ref{fig:pca_ssh}) for the spin-state data fail to reveal any distinct features for a wide range of couplings across the transition, whereas the spin-opstring data exhibits a clear trend from which even the phase transition can be determined (by the location of the trough). This indicates that spin-opstring carries information regarding the transition already in such a manner that can be detected by linear dimensionality reduction techniques like PCA.

However, non-linear methods like neural networks might still be able to extract necessary information from the spin-state. 
To illustrate the advantage of the combined spin-opstring data over the spin-state alone, using neural networks, we consider a system of HCBs on a 2D periodic honeycomb lattice governed by the Hamiltonian,
\begin{align}\label{Eq:Ham_WTI}
    H=-t_1\sum_{\langle i,j\rangle}(b_i^\dagger  b_j+\textrm{h.c.})+\sum_i W_i  n_i-\mu\sum_i n_i.
\end{align}
Here $t_1$ denotes the NN hopping amplitude and the on-site potential $W_i$ alternates between the values $W_0$ and $-W_0$ on even and odd $y-$layers respectively. The rest of the notation is already explained below Eq.~\ref{Eq:Hamiltonian}. This model exhibits a rich phase diagram, including a weak topological insulator (WTI), which has been investigated using SSE QMC \cite{10.21468/SciPostPhys.10.3.059} and recently, using ML networks trained on correlation function data extracted from SSE \cite{PhysRevB.110.165134}. For convenience, we call this the 2D WTI model.
 
We focus on the line at $W_0=0.5$ in the phase diagram, where as a function of the chemical potential $\mu$, the system goes through two phase transitions, from an empty phase at density $\rho=0$ to a superfluid phase, followed by a transition to a MI phase at density $\rho=1$. We have collected the training data from deep within the three phases and trained a ML model separately with 
spin-state and spin-opstring as input data types. The architecture of the ML model is same as the one used for predicting the phase boundaries of the $t_2-V_1$ model.

\begin{figure}[t]
    \centering
    \includegraphics[width=0.9\linewidth]{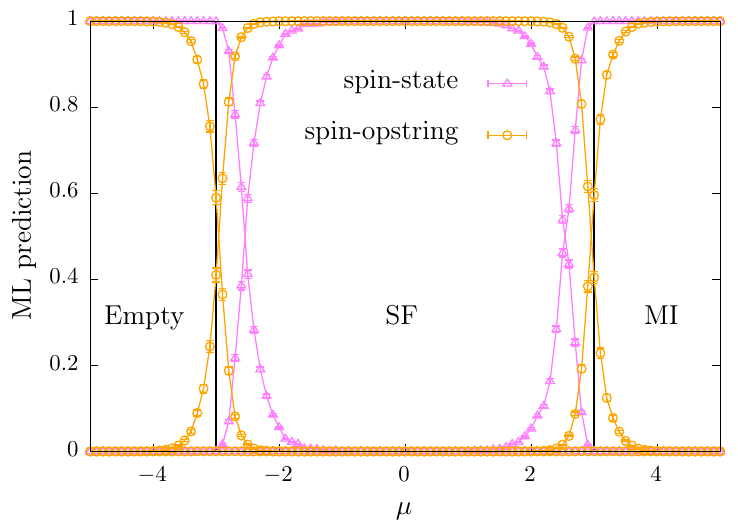}
    \caption{Predictions for the phase transitions in the 2D WTI model by ML models trained with spin-state and spin-opstring data. The black vertical lines indicate the true phase transition points.}
    \label{fig:wti_spinopstring}
\end{figure}

\begin{figure}[t]
    \centering
    \includegraphics[width=0.7\linewidth]{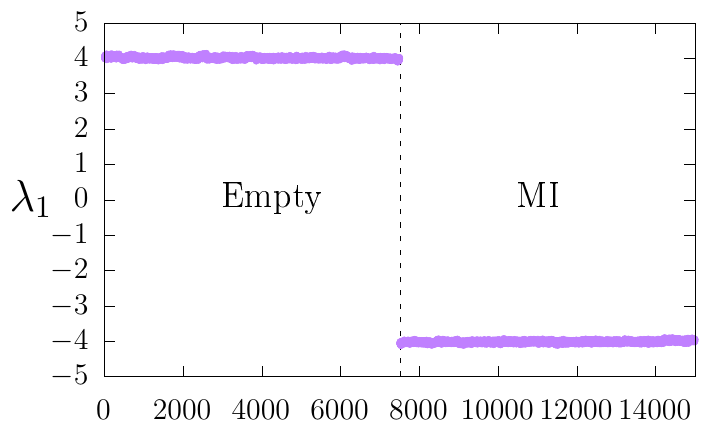}
    \includegraphics[width=0.7\linewidth]{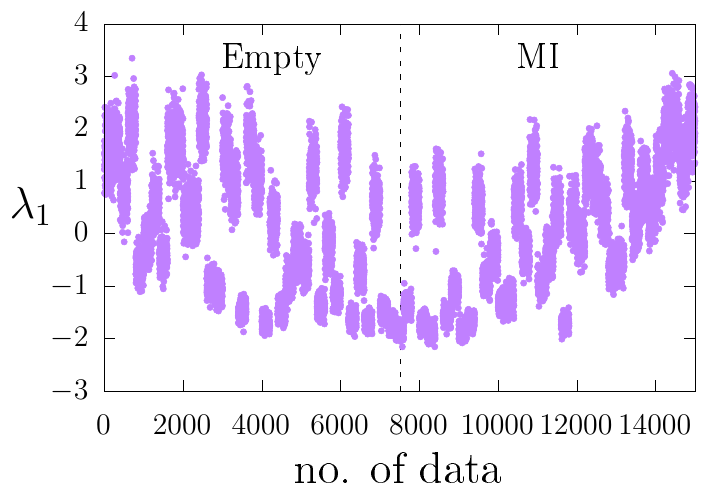}
    \caption{Variance $\lambda_1$ of the first principal component obtained from PCA done on data collected from deep within the empty phase and Mott insulator (MI) phase. The \textit{top} plot shows the result of the analysis on the combined spin-opstring data while the \textit{bottom} plot shows the case with only opstring data. }
    \label{fig:pca_t2v}
\end{figure}

When the ML model is trained with spin states only, the training and validation losses converge and decrease with increasing epoch number. However, when tested with unseen data across the two transitions, the performance of the model does not appear to meet the mark, as seen in Fig.~\ref{fig:wti_spinopstring}. Although it captures two phase transitions, the predicted transition points deviate from the actual ones by more than $16\%$. The ML model underestimates the width of the superfluid phase in terms of chemical potential. The reason for this poor performance is likely due to the fact that the superfluid phase persists over a wide range of densities, ranging from $0$ to $1$ with a large variation in the superfluid density. Since the identification of superfluidity crucially depends on information along the imaginary time direction, spin states alone do not carry enough essential information for this classification task.

The above result is, in fact, consistent with the conclusions in Ref.~\cite{PhysRevB.110.165134}. There it was observed that a ML model trained on density-density correlation function, which is a static quantity obtained from the spin-state, was unable to correctly predict the above transitions. On the other hand, the spatial correlation function, which encodes information of the temporal direction of the simulation cell, performed comparatively better\footnote{It should be noted that the data for spatial correlation function in the empty and MI phases are related by particle-hole symmetry, similar to the opstring. Therefore, the prediction for the empty-SF and SF-MI transitions were performed separately with two separately trained ML models.}. 

In contrast, when trained with spin-opstring data which encode both spatial and temporal information, the ML model captures the transition points accurately as can be seen in Fig.~\ref{fig:wti_spinopstring}. 
We have rigorously tested the ML models across 100 independent training runs with random initializations. For every run, models trained on spin-opstring data consistently predict the phase transitions accurately, while those trained on spin-state fail to do so. Thus, our findings demonstrate the advantage of spin-opstring data over spin-state alone, establishing it as essential for accurate and reliable identification of phase transitions in general.

\begin{figure*}
    \centering
    \includegraphics[width=0.8\linewidth]{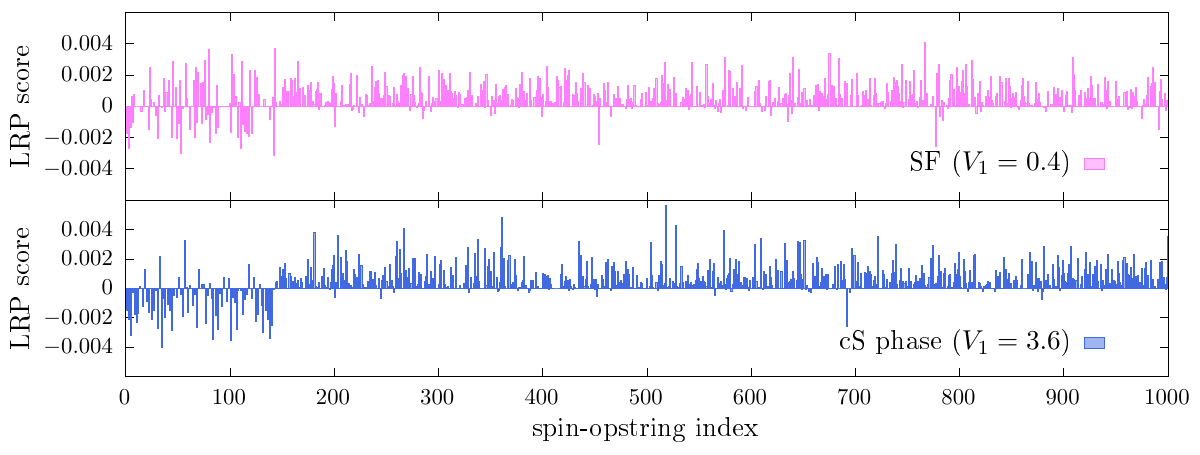}
    \caption{Relevance scores obtained using LRP-$\epsilon$ method for test data points in deep in the SF and cS phases. The first 1000 entries out of 17445 elements have been plotted for better visualisation. The first 144 indices correspond to the spinstate of a $12\times 12$ lattice.}
    \label{fig:LRP_map}
\end{figure*}

Furthermore, it would also be interesting to ask whether the opstring alone may suffice for ML tasks.
In fact, we find that when trained solely on the opstring data, a ML model is capable of predicting the phase transitions of the 2D $t_2-V_1$ model, discussed above in Sec.~\ref{sec:2dt2v}.
But, the opstring may not be effective as a standalone training input between two phases which are related by some symmetry.
For example, in the 2D $t_2-V_1$ model, the empty phase at $\rho=0$ and the Mott insulator at $\rho=1$, which are related by a particle-hole symmetry transformation, have similar QMC operator strings but very different spin-states. To illustrate this, in Fig.~\ref{fig:pca_t2v} we show the variance along first principal component from PCA performed separately on spin-opstring and opstring data, collected deep within the empty and the MI phases. While the distinction between the combined spin-opstring in the two phases is evident, the operator strings exhibit similar features in both. This lack of distinction is further illustrated using neural networks in the following discussion.

In the 2D WTI model, we train the same ML model now with the opstring data alone to detect the phase transitions at $W_0=0.5$.
During the training process, we observe that the training loss decreases to $\mathcal{O}(10^{-5})$, while the validation loss keeps growing, which is a classic sign of overfitting. The reason behind this can be understood in the following way. As the empty and MI phases are related by particle-hole symmetry, the opstring for these two phases are very similar.
Therefore, when trained with these almost identical opstring data with different phase labels, the ML model attempts to minimize the loss on the training set by memorizing the contradictory labels.
This prevents it from learning generalizable patterns, causing its performance to degrade on unseen validation data. 

Thus, our study shows that the combination of the spinstate and opstring proves effective in both avoiding data duplication and encoding quantum information necessary for the classification task. This brings us to the conclusion that while spin state or opstring data alone may suffice in some cases, the spin-opstring input proves to be more reliable and broadly applicable across different scenarios.

\section{Interpretability}\label{interpretability}

Despite the impressive prediction accuracy of the ML models, their inherent ``black-box" nature can obscure the physical principles driving their decisions.  
Model interpretability techniques allow us to ensure if the network is making decisions based on physically meaningful features, 
rather than exploiting artifacts or biases in the data. Following the demonstration of the broad applicability of the spin-opstring data across various ML tasks, the next logical step would be to understand what do the ML models learn from this data type.

\begin{figure}
    \centering
    \includegraphics[width=0.8\linewidth]{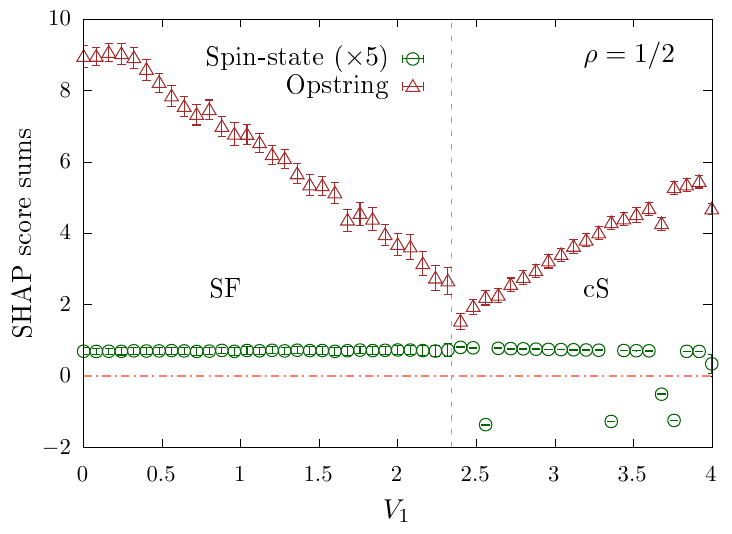}
    \includegraphics[width=0.8\linewidth]{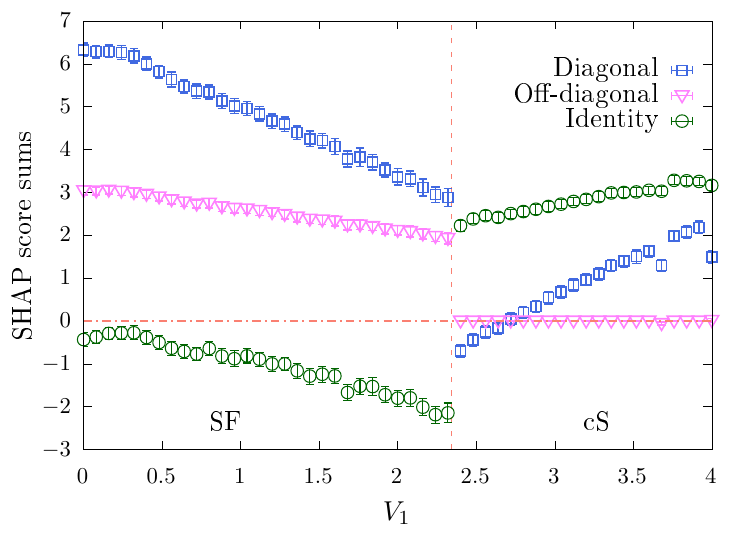}
    \caption{SHAP analysis for the SF-cS transition  as a function of increasing repulsion $V_1$. Top: Score sums for spin-state (scaled for clarity) and operator string. Bottom: Score sums of individual operators. Predicted transition point is shown as the dashed vertical line.}
    \label{fig:VT_SHAP}
\end{figure}

\begin{figure}
    \centering    \includegraphics[width=0.8\linewidth]{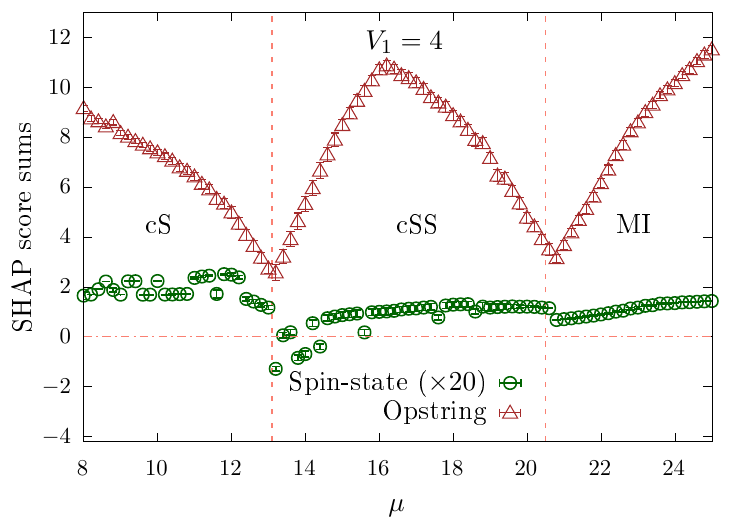}
    \includegraphics[width=0.8\linewidth]{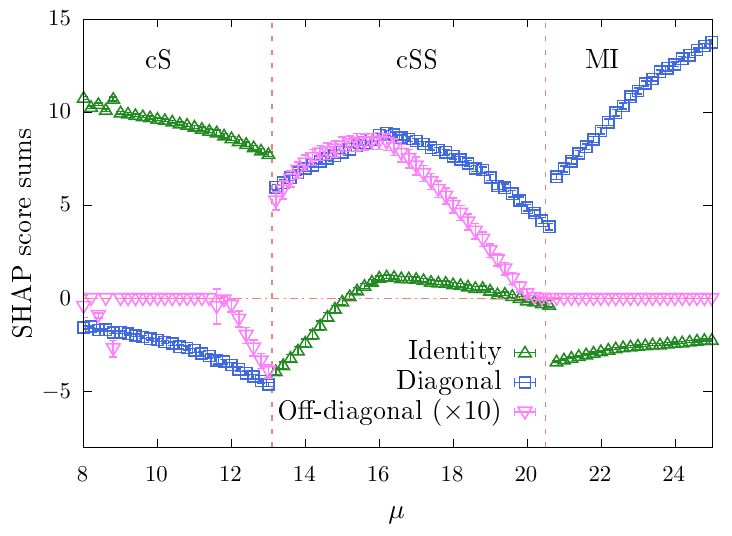}
    \caption{SHAP analysis of the cS–cSS–MI transition as a function of chemical potential ($\mu$) with fixed repulsion ($V_1=4.0$). Top: Score sums for spin state (scaled for clarity) and the operator string. Bottom: Contributions for different operator types; off-diagonal sum has been scaled by a factor for better comparison. The dashed vertical lines represent predicted transition points.}
    \label{fig:HT_SHAP}
\end{figure}

A key advantage of the spin-opstring representation is that its structured format naturally supports interpretability of ML predictions based on physical phenomena. The data consists of two components: the spin configuration and the operator string, which includes diagonal, off-diagonal, and identity operators, each associated with specific physical processes. This structure allows us to examine which features the model prioritizes in different phases. For example, in superfluid phase, the model focuses on off-diagonal operators that capture particle hopping, whereas in insulating phases, diagonal and identity operators become more relevant. Notably, in the 1D SSH model, where a local order parameter is absent, the network distinguishes the two dimerized phases through an inverse relevance pattern between off-diagonal and identity operators, which directly reflects the alternating bond structure central to SSH physics. Overall, the underlying structure of the spin-opstring enables meaningful insights into the model’s behavior, something not easily achieved with conventional inputs.

In this work, we employ two widely used  interpretability
techniques, Layer-wise Relevance Propagation (LRP) and SHapley Additive exPlanations (SHAP). Both these methods aim to explain a trained ML model's prediction by distributing the final neuron outputs back to the input features, thereby quantifying each feature's contribution. For example, in Fig.~\ref{fig:LRP_map} we show the raw relevance or scores for input features using LRP, obtained for test data points within the SF and cS phases in the 2D $t_2-V_1$ model. Unlike relevance map visualizations
of images like handwritten digits, a direct interpretation of individual features of the spin-opstring appears challenging. To extract physically meaningful insights, we leverage the underlying structure to compute separate sums of the LRP or SHAP scores for spin-state, opstring and each of the three individual operator types.

We observed a key difference in how each method attributes relevance. LRP, due to its reliance on input values for relevance propagation, assigns zero relevance to identity operators, which are encoded as zeros in the spin-opstring data. In contrast, SHAP can assign non-zero relevance to identity operators if their presence or absence influences the model’s output. Since identity operators are an integral part of our spin-opstring data, SHAP appears to be more suitable than LRP for our analysis. Therefore, in the main text, we present our interpretability results using SHAP, while details of the LRP-based analysis are included in Appendix~\ref{app4}. Apart from the discrepancy regarding identity operators, the LRP results support the SHAP analysis presented below.

SHAP is a model-independent method, based on cooperative game theory, where each input feature is treated as a ``player" in a game \cite{NIPS2017_7062}. It aims to fairly distribute the model's prediction among the features  
by measuring how the model's output changes when the input features are systematically added or removed.
In its general form (\textit{e.g.}, KernelSHAP), SHAP is model-agnostic, which makes it particularly valuable for validating interpretability results across a wide range of models. 
In our study, we utilize the Deep SHAP algorithm, a variant tailored for deep neural networks, to compute the SHAP scores, using the Python library based on \cite{NIPS2017_7062}.

We start our interpretability analysis with the phase classification tasks for the 2D $t_2-V_1$ model. 
The trained ML models used for the prediction of various phase transitions is analysed on the corresponding test datasets. For better interpretability, the ML model without the final softmax activation layer is utilised for this purpose and we consider results  with respect to the output neuron having the highest logit value. Throughout this section, the error bars represent the standard deviation of interpretability scores for a single model across multiple test samples at each coupling value.

The total SHAP scores for the spin-state and opstring, shown in Fig.~\ref{fig:VT_SHAP} top as a function of $V_1$, reveal that the model primarily uses the opstring for SF-cS phase classification. Since the spin-state scores are nearly identical across both phases, it confirms that the model doesn't rely on this feature. As depicted in Fig.~\ref{fig:VT_SHAP} bottom, a closer look at the operator types shows that all of them exhibit clear jumps at the transition point. Diagonal operators are relevant in both phases, while off-diagonal operators contribute only in the SF phase, consistent with the presence of particle hopping. Within the insulating cS phase, identity operators become the dominant contributors to the model's prediction.

The SHAP analysis for the cS-cSS-MI transition, reveals that as in the SF-cS case, the contribution of the spin-state is much smaller than the opstring (Fig.~\ref{fig:HT_SHAP} top). Consistent with their insulating nature, the cS and MI phases show almost no relevance for off-diagonal operators as depicted in Fig.~\ref{fig:HT_SHAP} bottom. In the cSS phase, however, which
is a homogeneous coexistence of superfluidity and cS,
both the diagonal and off-diagonal operators become important. Diagonal operators are also relevant in
the MI phase and the increasing score sums with $\mu$ are actually proportional to the number of diagonal
operators present in the opstring. Consistent with the cS-SF transition, identity operators contribute the most in the identification of the cS phase.  This provides a coherent and physically meaningful picture of how the model identifies a phase.

\begin{figure}
    \centering
    \includegraphics[width=0.8\linewidth]{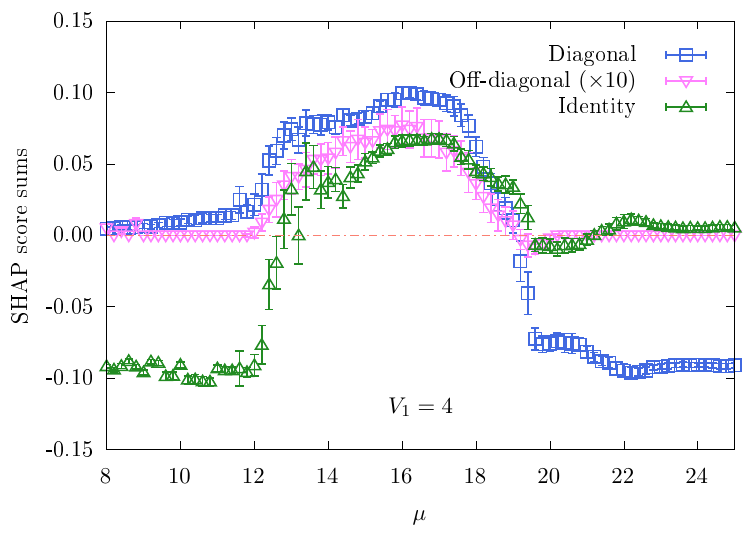}
    \caption{Summed SHAP scores for different operator types for the regression task of predicting superfluid density across the cS–cSS–MI phase transition in the 2D $t_2-V_1$ model with varying chemical potential $\mu$.}
    \label{fig:reg_SHAP}
\end{figure}

Next, we performed SHAP analysis for the regression task in the cS-cSS-MI phase transitions with superfluid density $\rho_s$ as the regression target (cf.~Fig.~\ref{fig:HT_reg}).  
As the spin-state contribution is much smaller than the opstring, we focus on the individual operator types in Fig.~\ref{fig:reg_SHAP}.
For the cSS phase, which has a finite $\rho_s$ value, all operator types contribute positively, which is consistent with the classification results (Fig.~\ref{fig:HT_SHAP} bottom).
However, for the cS and MI phases, SHAP assigns negative values for the key features identified in the classification task, \textit{i.e.,} identity operators for the cS phase and diagonal operators for the MI phase.

This discrepancy arises from the fundamentally different objectives of classification and regression.  
Since the cS and MI phases both exhibit zero $\rho_s$, features characteristic of these phases act to push the predicted $\rho_s$ value towards zero.  
Given a positive average prediction baseline set by the cSS data, 
features that reduce the predicted value below this baseline naturally receive negative SHAP values.
Thus, the negative scores of identity and diagonal operators in the cS and MI phase, respectively, do not imply irrelevance, but rather reflect their effectiveness in suppressing the predicted $\rho_s$ to zero.

\begin{figure}
    \centering
    \includegraphics[width=0.8\linewidth]
    {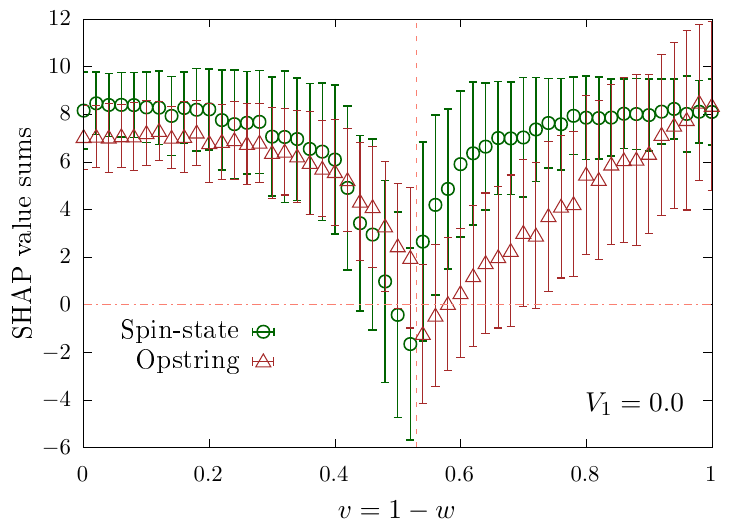}
    \includegraphics[width=0.8\linewidth]{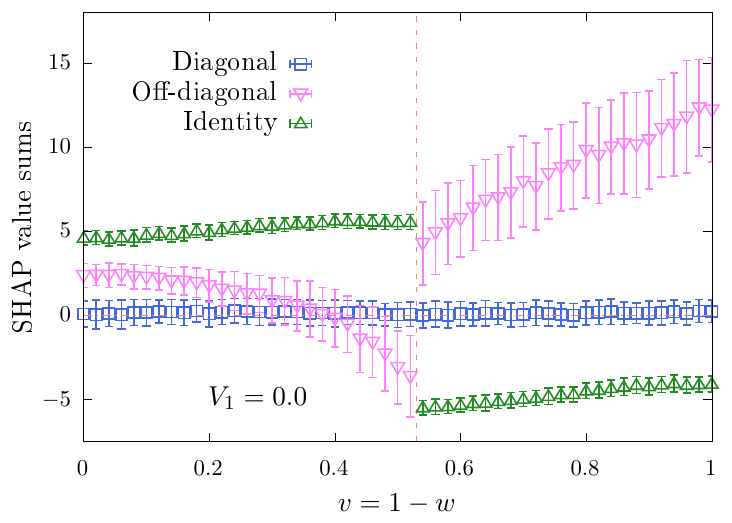}
    \includegraphics[width=0.8\linewidth]{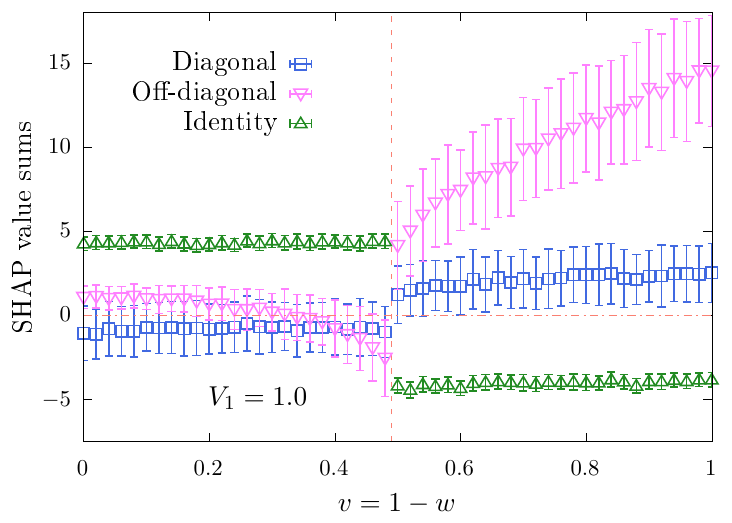}
    \caption{SHAP analysis for non-interacting and interacting SSH model. Top: SHAP value sums for the spin state and operator string across the topological–to-trivial phase transition in the non-interacting SSH model. Middle: SHAP value sums for individual operator types in the non-interacting case. Bottom: Summed SHAP values for operators in the interacting SSH model with nearest-neighbor repulsion $V_1=1.0$. The vertical dashed lines in all plots indicate the predicted phase transition point.}
    \label{fig:SHAP_SSH}
\end{figure}

In Sec.~\ref{trans_learning} we demonstrated the performance of the spin-opstring in a transfer learning setting, where a ML model was trained on labelled data from the non-interacting SSH model and evaluated on its interacting version. To understand what features the machine learns in both regimes, we have applied SHAP to the non-interacting model as well as the interacting model (with $V_1=1.0$). 
In Fig.~\ref{fig:SHAP_SSH} top, the SHAP value sums show that the spin-state and opstring contribute almost equally to the predictions in the non-interacting case. Fig.~\ref{fig:SHAP_SSH} middle further breaks down opstring contributions by operator type, revealing that diagonal operators carry zero SHAP value, while identity and off-diagonal operators display an inverse relevance pattern. Specifically, when one is positively relevant for a phase, the other consistently shows low or negative relevance, and vice-versa for the alternate phase.

These observations offer a crucial insight into how the ML model distinguishes the phases of the non-interacting SSH model. Diagonal operators encode static properties which do not vary significantly across the transition, 
so, the ML model does not rely on them to identify either of the two phases. 
On the other hand, off-diagonal operators represent hopping,  
while identity operators signify its absence. The complementary behavior of these two operators in the SHAP analysis indicates that the machine has learned the SSH model's dimerization pattern, \textit{i.e.}, the alternating strong and weak bonds. For example, if a phase features strong hopping on even bonds, the model assigns high importance to the off-diagonal operators at those bonds.  
However, identity operators on even bonds makes the model less confident in that phase, resulting in their negative relevance.
The opposite pattern holds for the other phase. This clear connection between feature importance and the underlying physical phenomenon shows that the model is not merely fitting the data, it is learning the dimerization pattern that distinguishes the topological and trivial phases of the SSH model.
 
Introducing interactions in the SSH model primarily affects the relevance of diagonal operators, while the importance of other operators remains largely unchanged. In Fig.~\ref{fig:SHAP_SSH} bottom, the summed SHAP values for the interacting case ($V_1=1.0$) indeed reveal that the diagonal operators now have nonzero relevance, positive in one phase and negative in the other. 
This important finding shows that the ML model, with its trained knowledge of the non-interacting system, can utilise features in the interacting Hamiltonian. 
Although the model never interpreted diagonal operators as meaningful during training, it correctly identifies their role during inference on interacting data, a hallmark of effective generalization in transfer learning. This behavior underscores that spin-opstring encodes the underlying physics in a way that allows a model to adapt.
  
A recent interesting work on the 1D disordered SSH model \cite{Cybinski_2025} employed another interpretability technique, class activation mapping, to gain insights into trained CNN models. 
Our study offers a complementary perspective in the interacting SSH model with a different input data, the spin-opstring, which allows successful interpretability using simple ML models.

\section{Conclusions}\label{conclusion}
In conclusion, we have proposed to use the ``spin-opstring", extracted from Stochastic Series Expansion QMC as a input data type for machine learning applications. The spin-opstring is a compact representation of the QMC simulation cell that retains all quantum information of the system while being memory-efficient. It consists of the initial state of the simulation cell followed by an operator string that describes the state's evolution through imaginary time, ultimately bringing it back to its original configuration. 
To demonstrate the applicability of spin-opstring as ML input, we consider two different models : the 2D $t_2-V_1$ model and the 1D SSH model. We find that trained ML models are able to successfully predict conventional and topological phase transitions. In a regression setting, ML models were also able to correctly predict the superfluid density in the 2D $t_2-V_1$ model.
We have further compared the performance of the spin-opstring with the full simulation cell data. Our results show that as system size increases, the combined memory and computational costs make the full simulation cell impractical.

We showcase broader applicability of spin-opstring with successful transfer learning experiments. First, we show that a ML network trained on the non-interacting SSH model accurately predicted phase transitions in its interacting version up to moderate interaction strength. In another example, a model trained on the $t_2-V_1$ system correctly predicted a similar phase transition in the $t_1-V_1$ system. These results illustrate the strong potential of spin-opstring data in facilitating knowledge transfer between related models. Furthermore, by training a ML network on a smaller lattice and predicting transitions on a larger system, we confirm that spin-opstring supports generalization across system sizes.

Overall, our results establish that spin-opstring can successfully be applied across a wide range of ML tasks.
However, this raises the question of whether it can surpass conventional inputs based on spin configurations. In a HCB system, dubbed as the 2D WTI model, we demonstrate the superiority of spin-opstring over spin-state as ML input in accurately predicting a quantum phase transition. We also present a case where using the opstring alone as input causes training issues. These findings demonstrate that the unified spin-opstring representation leads to more reliable model performance.

Finally, we employed two leading interpretability techniques, Layer-wise Relevance Propagation (LRP) and SHapley Additive exPlanations (SHAP), to gain insight into the model’s decision-making processes. Leveraging the internal structure of spin-opstring, we devised a strategy to make the high-dimensional outputs of LRP and SHAP interpretable by linking the feature scores to specific physical operators. Our analysis revealed that while LRP disregards identity operators (as they are encoded as zeros in the spin-opstring), SHAP is able to recognize their contextual importance. This highlights the advantage of SHAP in scenarios where every component of the input, including identity operators, contribute to the model’s predictions. Across classification, regression, and transfer learning tasks, our interpretability study consistently showed that the models learned physically meaningful features from the input data. This supports the conclusion that the spin-opstring is not only effective but also interpretable for ML-based analysis of quantum systems.

In summary, our results establish the spin-opstring as a broadly applicable, generalizable, and interpretable input format for machine learning in quantum many-body physics.
It would be interesting to see how the spin-opstring data perform in the domain of unsupervised ML algorithms. Another interesting direction of our future research involves the use of generative ML models with the spin-opstring data representation for the generation of SSE configurations.

\begin{acknowledgments}
    This work was supported by the Center for Theory and Computation, National Tsing Hua University-113H0001L9, Hsinchu 300 Taiwan, the Taiwan National Science and Technology Council Project  No.~112-2639-M-002-006-ASP and NSTC 113-2112-M-002-033-MY3.
\end{acknowledgments}

\appendix
\section{Effect of Batch Normalization}\label{app1}
\begin{figure}[t]
    \centering
    \includegraphics[width=0.8\linewidth]{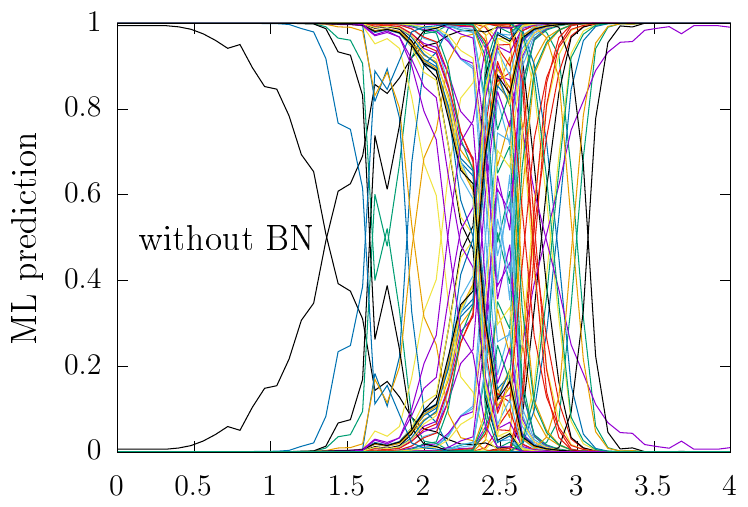}
    \includegraphics[width=0.8\linewidth]{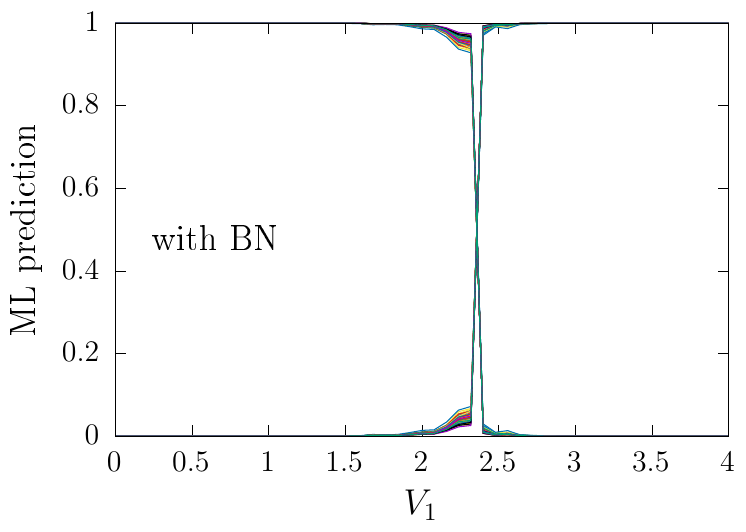}
    \caption{Effect of Batch Normalization (BN) on machine prediction for the SF-cS transition in the 2D $t_2-V_1$ model at half-filling, as a function of NN repulsion $V_1$. \textit{Top}: Predictions from $100$ independently trained models without BN. \textit{Bottom}: The same number of independent predictions from the ML model with added BN layers.}
    \label{fig:batchnorm}
\end{figure}

Batch Normalization (BN) is a well-known method in neural network machine learning which helps to train the models faster and also acts as a regularization to prevent overfitting \cite{ioffe2015batch}. It acts by first``whitening'' \textit{i.e.}, transforming linearly to have zero mean and unit variance, the outputs of a network layer and further scale and shift the resulting output with trainable parameters. This results in better convergence of the gradient descent training process. In our study of phase classification, we have found that BN plays an important role in stabilizing the predictions of the phase transition points.

In Fig.~\ref{fig:batchnorm}, we compare the machine predictions for the SF-cS transition in the 2D $t_2 - V_1$ model at half-filling, as a function of NN repulsion $V_1$, obtained by training the same network with and without BN layers.
We show the predictions for $100$ independently trained models in each case using the same hyperparameters. The figure clearly shows that the ML model gives consistent predictions for the phase transition point when BN is applied. Although the predictions are correct deep in the phases, the predictions without BN show a wide variation in the predicted phase transition point. 

BN layers are commonly applied to ML networks to mitigate ``internal covariate shift'', a phenomenon where the distribution of inputs to a layer in a neural network changes during training due to the continuous updates of parameters in preceding layers. In our spin-opstring data derived from SSE QMC simulations, the operator string segments may exhibit considerable lengths (e.g., up to $5\times 10^4$ elements for $12\times 12$ lattices in the $2D$ $t_2-V_1$ model) and substantial fluctuations for each element, $S(p)$. As argued in Ref.~\cite{ioffe2015batch}, the absence of BN in such scenarios can lead the ML model to overfit and be susceptible to the initial conditions of the training process. This is consistent with our observation of a wide range of predictions for the phase transition point in Fig.~\ref{fig:batchnorm} \textit{top}.

\section{Domain adaption from $t_2-V_1$ to $t_1-V_1$}\label{app2}

To explore the application of transfer learning we consider another example; the well-known $t_1-V_1$ model on the periodic square lattice, governed by the Hamiltonian,
\begin{align}
 H=-t_1\sum_{\langle i,j\rangle}(b^\dagger_i b_j+\textrm{h.c.}) 
    +V_1\sum_{\langle i,j\rangle}n_in_j-\mu\sum_i n_i.
    \label{Eq:t1V1Hamiltonian}
\end{align}
Here, $t_1$ and $V_1$ denote hopping amplitude and repulsion, respectively, on the NN bonds. The reason for choosing this model lies in the similarity with phases in the $t_2-V_1$ model. Notably, both models depict phase transition from a superfluid phase (SF) to a checkerboard solid (cS) phase as a function of increasing NN repulsion $V_1$, at half-filling.

\begin{figure}[t]
    \centering
    \includegraphics[width=0.8\linewidth]{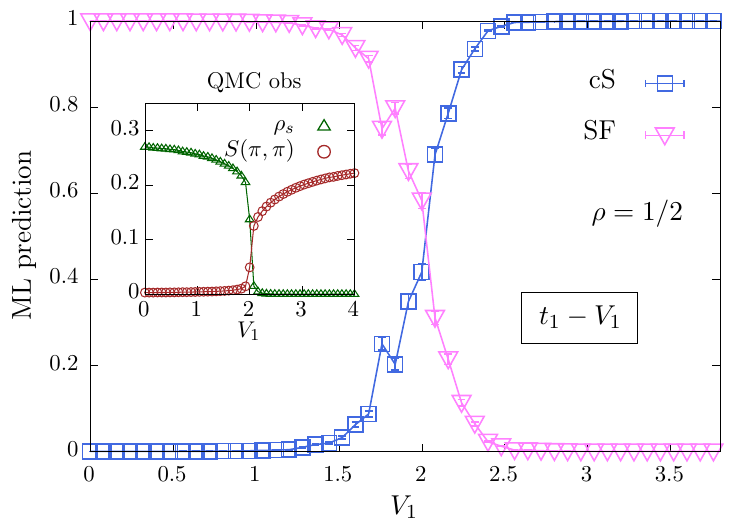}
    \caption{Prediction of the SF-cS phase transition in 2D $t_1-V_1$ model by a ML network trained on 2D $t_2-V_1$ model. \textit{Inset}: Superfluid density ($\rho_s$) and structure factor $S(\pi,\pi)$ obtained from SSE QMC simulations across the SF-cS transition in the $t_1-V_1$ model.}
    \label{fig:t1v}
\end{figure}

The ML model used for this transfer learning is trained with data sampled from deep within the cS and SF phases of the $t_2-V_1$ model and then tested on the prediction data collected across the SF-cS transition in the $t_1-V_1$ model. The data from both models on $12\times 12$ lattices are of similar dimensions and have been padded with zeros as described in Sec.~\ref{method}.

The prediction of this ML model when tested on the $t_1-V_1$ model is shown in Fig.~\ref{fig:t1v}. It depicts a clear phase transition from the SF to cS phase around $V_1=2.0$. To compare the prediction with the actual QMC results, we have plotted the superfluid density $\rho_s$ along with the structure factor $S(\pi,\pi)$ in the inset of Fig.~\ref{fig:t1v}. The transition point predicted by the ML model is indeed very close to the QMC results. This result is particularly noteworthy because of the following reason. Although both systems show a SF-cS transition, the microscopic origin of the SF phase in the two models are different. In the $t_1-V_1$ model the superfluidity arises from the hopping of bosons along the NN bonds, whereas in the $t_2-V_1$ model, it arises due to NNN hopping. 
Now, hopping of a particle along a bond $b$ corresponds to an off-diagonal operator labeled $2b+1$ in the opstring.
In our convention, NN bonds are numbered from $0$ to $2N_s$, whereas the numbering of NNN bonds lie between $2N_s+1$ to $4N_s$. Therefore the actual values arising in the opstring data in the SF phase of the two systems are completely different. It is interesting to observe that regardless of the microscopic details, the machine is able to learn the underlying feature of the fundamental nature of a phase.

\section{Details of architecture used in generalization across system sizes}\label{app3}

A possible way to perform the training for this task would be to equalize the spin-opstring lengths of the larger and smaller systems through padding. However, this method is not compatible with arbitrary length data during inference and it also has an additional computational overhead.
Instead, to handle the variable length data, we have incorporated the Global Pooling Layer in 1D CNN architecture. In particular for the 1D SSH model we keep the same network structure as before but replace the flattening layer with a global average pooling layer.
This layer reduces output of each CNN filter to a single value, which is an average over all its entries, thus effectively converting inputs of varying lengths into fixed-size output vectors. The subsequent FC layers always receive inputs of consistent size, allowing the model to generalize across different system sizes without needing padding or truncation.

Global pooling layers, such as global average or max pooling, provide a simple yet effective way to deal with variable-length inputs. They promote scale invariance, which is beneficial when the exact position of features is not essential. However, they can sometimes discard important spatial or sequential details, which might be crucial for certain tasks. Alternative approaches, like recurrent neural networks and attention mechanisms, can preserve this information, but tend to be more complex and computationally demanding.

\section{Interpretability analysis using Layer-wise Relevance
Propagation}\label{app4}

\begin{figure}
    \centering
    \includegraphics[width=0.8\linewidth]
    {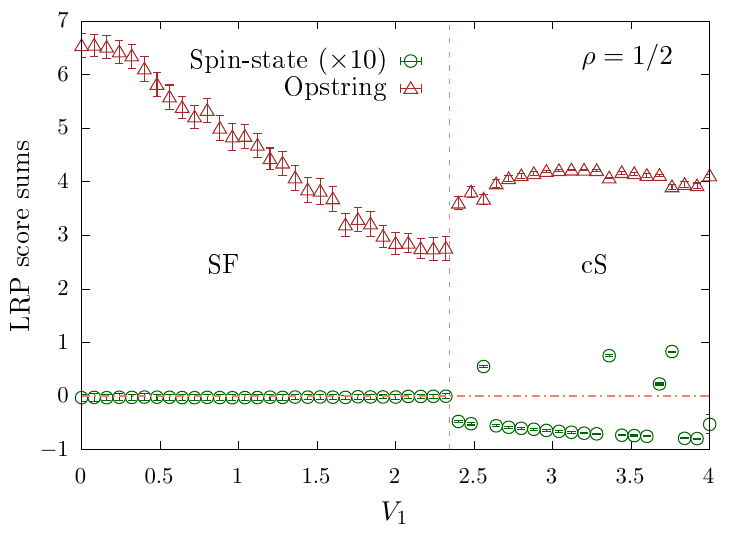}
    \includegraphics[width=0.8\linewidth]{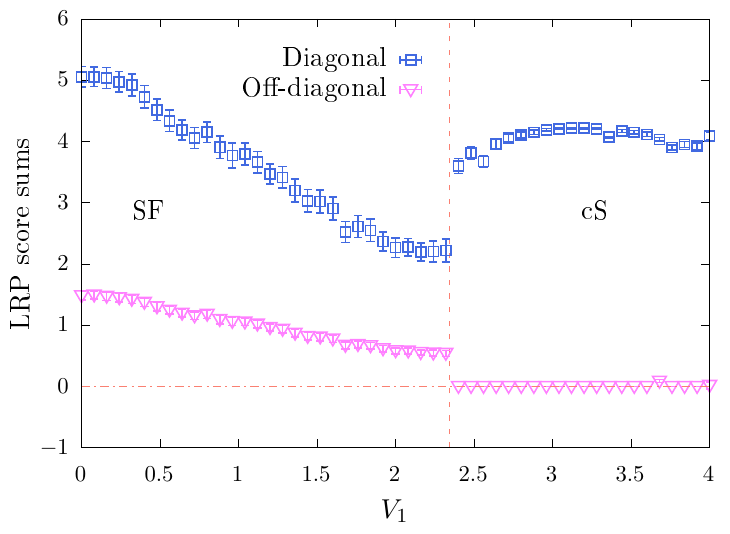}
    \caption{LRP analysis of the SF-cS transition in the 2D $t_2-V_1$ model. Top: Sum of LRP values for the spin-state (scaled for better visibility) and the operator string as a function of NN repulsion $V_1$.  
    Bottom: LRP sums for diagonal and off-diagonal operators, excluding identity operators with zero relevance. The dashed vertical lines in both plots indicate the predicted transition point.}
    \label{fig:LRP_VT}
\end{figure}

\begin{figure}
    \centering
    \includegraphics[width=0.8\linewidth]
    {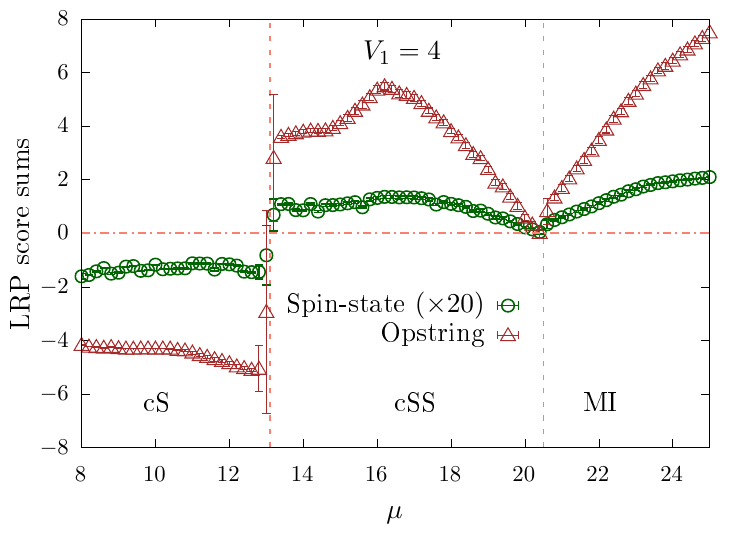}
    \includegraphics[width=0.8\linewidth]{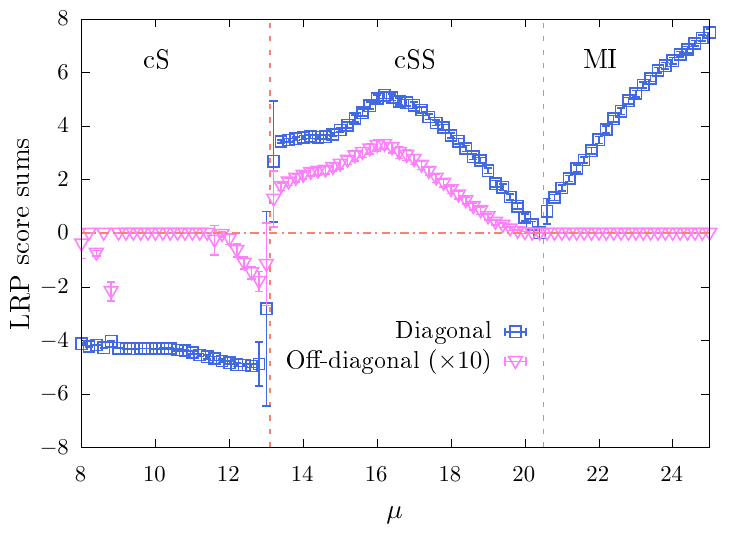}
    \caption{LRP analysis of the cS-cSS-MI transition as a function of chemical potential $\mu$ with repulsion fixed at $V_1=4.0$. Top: LRP sums for the spin-state (scaled for clarity) and the operator string. Bottom: LRP contributions from diagonal and off-diagonal operators; identity operators are omitted due to zero relevance. The spin-state and off-diagonal sums are scaled for better visibility. In both plots, the predicted transition points are indicated by dashed vertical lines.}
    \label{fig:LRP_HT}
\end{figure}

Layer-wise Relevance Propagation (LRP) is a powerful post hoc interpretability technique that decomposes the prediction of the model by propagating relevance backward from each output neuron to the input features \cite{10.1371/journal.pone.0130140}. With the application of a set of conservation-based propagation rules at each layer, LRP yields a relevance map over the input data.

For our LRP analysis, we use the Epsilon rule (LRP-$\epsilon$), where the propagation of relevance scores between two consecutive layers of the ML network is governed by,
\begin{equation}
    R_j = \sum_k \frac{z_{jk}}{\epsilon + \sum_j z_{jk}}R_k \,,
\end{equation}
where $R_j$ corresponds to the score for a neuron $j$ in the inner layer and $R_k$ for a neuron $k$ in the outer layer. The coefficients $z_{jk}$ represent the model weights and biases present in between the two layers.
The core idea of the LRP-$\epsilon$ rule involves adding a small positive constant $\epsilon$ to the denominator, the role of which is to offset some relevance from weak or contradictory neuron activations. Thus it helps to retain the most salient relevances by suppressing noise and avoids numerical instabilities arising from vanishing activations. We have used $\epsilon=10^{-3}$ for the LRP-$\epsilon$ rule and performed our LRP analysis using the \textit{iNNvestigate} library \cite{JMLR:v20:18-540} which allows integration with Keras.

We analyse the phase classification networks in the 2D $t_2-V_1$ model using LRP. Similar to SHAP, we use ML classifiers without the final softmax activation for the analysis. The interpretability results obtained from LRP are mostly consistent with SHAP with one major difference, LRP assigns zero relevance to identity operators and is therefore unable to ascertain their role. Consequently, we omit identity operators in the LRP figures below.

In Fig.~\ref{fig:LRP_VT}, we show the total LRP sums for spin-state and opstring (top), and the individual operator types (bottom) across the SF–cS transition.
Consistent with the SHAP analysis, the spin-state is far less important for this classification than the opstring. The spin-state's  relevance drops to zero in the SF phase, likely due to the absence of any spin-order.
In the cS phase, however, its positive and negative relevances correspond to the two degenerate spin configurations.
The LRP scores for diagonal and off-diagonal operators show a qualitatively similar trend to SHAP results. However, as mentioned earlier, LRP is unable to identify the role of identity operators in the cS phase.

Next, for the cS-cSS-MI phase transition, we show the LRP relevance scores for the spin-opstring components in Fig.~\ref{fig:HT_12x12}. In the cSS and MI phases, the score trends for spin-state, opstring and different operators are similar to those from the SHAP analysis. 
In the cS phase, however, a discrepancy arises, the total LRP sum for the spin-opstring is negative, which contradicts the positive logit value from the model's output. This result appears to be a significant violation of the LRP conservation property.

The source of this discrepancy  becomes clear when we
compare the LRP analysis with the SHAP results (Fig.~\ref{fig:HT_SHAP}). In the cS phase, both methods assign negative scores to diagonal operators, while off-diagonal ones remain zero. The total SHAP sum, however, becomes positive due to a large positive contribution from identity operators. Since LRP assigns zero relevance to these operators, its total score fails to match the positive logit value.
This contradiction originates from a fundamental difference in how the two methods attribute relevance. In the spin-opstring data, identity operators are encoded as zero. Since LRP propagates relevance based on neuron activations and weights, it tends to assign zero relevance to features that are numerically zero in the input \cite{samek2016interpreting}, regardless of their contextual role. In contrast, SHAP assigns non-zero relevance to identity operators by measuring how its presence or absence influences the model’s prediction.

\bibliography{ML_simcell}
\end{document}